\documentclass[manuscript,screen]{acmart}

\AtBeginDocument{%
  \providecommand\BibTeX{{%
    \normalfont B\kern-0.5em{\scshape i\kern-0.25em b}\kern-0.8em\TeX}}}

\usepackage{subfig}
\usepackage{multirow}

\setcopyright{acmcopyright}
\copyrightyear{2018}
\acmYear{2018}
\acmDOI{XXXXXXX.XXXXXXX}

\acmConference[Conference acronym 'XX]{Make sure to enter the correct
  conference title from your rights confirmation emai}{June 03--05,
  2018}{Woodstock, NY}
\acmPrice{15.00}
\acmISBN{978-1-4503-XXXX-X/18/06}

\begin{document}

\title{\textit{SwitchX}: Gmin-Gmax Switching for Energy-Efficient and Robust Implementation of Binarized Neural Networks on ReRAM Xbars}

\author{Abhiroop Bhattacharjee}
\affiliation{%
  \institution{Yale University}
  \city{New Haven}
  \country{USA}
  }
\email{abhiroop.bhattacharjee@yale.edu}

\author{Priyadarshini Panda}
\affiliation{%
  \institution{Yale University}
  \city{New Haven}
  \country{USA}
  }
\email{priya.panda@yale.edu}

\renewcommand{\shortauthors}{Bhattacharjee, et al.}

\begin{abstract}
  Memristive crossbars can efficiently implement \textit{Binarized Neural Networks} (BNNs) wherein the weights are stored in high-resistance states (HRS) and low-resistance states (LRS) of the synapses. We propose \textit{SwitchX} mapping of BNN weights onto ReRAM crossbars such that 
the impact of crossbar non-idealities, that lead to degradation in computational accuracy, are minimized. Essentially, \textit{SwitchX} maps the binary weights in such manner that a crossbar instance comprises of more HRS than LRS synapses. 
We find BNNs mapped onto crossbars with \textit{SwitchX} to exhibit better robustness against adversarial attacks than the standard crossbar-mapped BNNs, the baseline. Finally, we combine \textit{SwitchX} with state-aware training (that further increases the feasibility of HRS states during weight mapping) to boost the robustness of a BNN on hardware. We find that this approach yields stronger defense against adversarial attacks than adversarial training, a state-of-the-art software defense. We perform experiments on a VGG16 BNN with benchmark datasets (CIFAR-10, CIFAR-100 \&  TinyImagenet) and use \textit{Fast Gradient Sign Method} ($\epsilon=0.05$ to $0.3$) and \textit{Projected Gradient Descent} ($\epsilon=\frac{2}{255}$ to $\frac{32}{255},~\alpha=\frac{2}{255}$) adversarial attacks. We show that \textit{SwitchX} combined with state-aware training can yield upto $\sim35\%$ improvements in clean accuracy and $\sim6-16\%$ in adversarial accuracies against conventional BNNs. Furthermore, an important by-product of \textit{SwitchX} mapping is increased crossbar power savings, owing to an increased proportion of HRS synapses, that is furthered with state-aware training. We obtain upto $\sim21-22\%$ savings in crossbar power consumption for state-aware trained BNN mapped via \textit{SwitchX} on 16$\times$16 \& 32$\times$32 crossbars using the CIFAR-10 \& CIFAR-100 datasets.
\end{abstract}



\keywords{Binarized neural network, ReRAM crossbar, non-idealities, switched-mapping, adversarial robustness}

\maketitle

\section{Introduction}

\label{sec:intro}
Memristive crossbars have received significant focus for their ability to realize \textit{Deep Neural Networks} (DNNs) by efficiently performing \textit{Multiply-and-Accumulate} (MAC) operations using analog dot-products ~\cite{schuman, mo-rram, sharad}. These systems have been realized using a wide range of emerging \textit{Non-Voltalile-Memory} (NVM) devices such as, \textit{Resistive RAM} (ReRAM), \textit{Phase Change Memory} (PCM), \textit{Ferroelectric FET} (FeFET) and Spintronic devices~\cite{chen, sengupta, sttsnn, mehonic2019simulation}. These devices exhibit high on-chip storage density, non-volatility, low leakage and low-voltage operation and thus, enable compact and energy-efficient implementation of DNNs~\cite{chakraborty2020pathways, puma, rxnn}. 

\begin{figure}[t]
    \centering
   \includegraphics[width=0.7\linewidth]{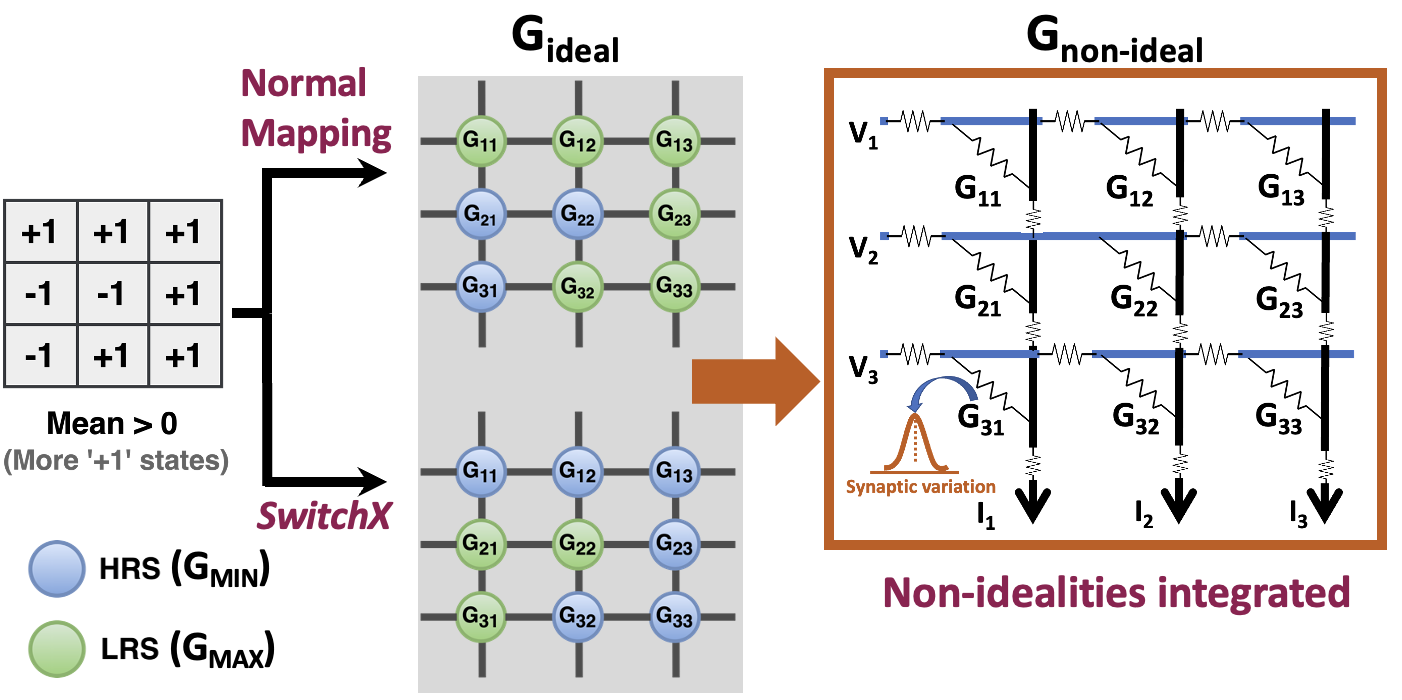}%
    \caption{Pictorial representation of Normal Mapping and \textit{SwitchX} approaches showing mapping of a 3x3 binary weight matrix on 3x3 crossbar with non-idealities. In the case of Normal Mapping, `+1's are mapped to LRS states and `-1's are mapped to HRS states to obtain $G_{ideal}$. In the case of \textit{SwitchX}, the mapping depends on the mean of the weight matrix. If mean $<$ 0, then the process is similar to Normal Mapping. If mean $>$ 0, then `+1's are mapped to HRS and `-1's to LRS. This ensures that HRS states are always in majority in a crossbar array.}
\label{mapping}
   
\end{figure}

In the recent years, \textit{Binarized Neural Networks} (BNNs) have emerged as efficient and reasonably accurate low-precision models to implement DNNs \cite{hubara2016binarized}. BNNs consist of binary synaptic weights and activations, \textit{viz.} \{-1,+1\}. The binarization enables lower computational complexity and power consumption for MAC operations. BNNs are crossbar-friendly in the sense that they can be implemented on crossbars in the manner shown in \figurename{~\ref{mapping}} (see Normal Mapping). Here, the binarized weights are programmed as conductances of the synaptic devices, such as ReRAMs, at the cross-points. A weight value of `+1' corresponds to a \textit{Low Resistance State} (LRS) while that of `-1' corresponds to a \textit{High Resistance State} (HRS). During inference, the activations of BNNs are fed into each row $i$ of the crossbar as analog voltages (generated using Digital-to-Analog Converters (DACs)). As per Ohm's Law, the voltages  interact with the device conductances at the cross-points $G_{ij}$ and produce a current. Consequently, by Kirchhoff's current law, the net output current sensed at each column $j$ is the sum of currents through each device, \textit{i.e.}:

\small
\begin{equation} \label{eq:dot-prd}
I_j = \sum_{i}^{}{G_{ij} * V_i}
\end{equation}
\normalsize

The above dot-product operation is depicted in \figurename{~\ref{xbar}} (Left). 

It is a known fact that memristive crossbar arrays possess non-idealities such as, interconnect parasitics, process variations in the synaptic devices, driver and sensing resistances, etc.~\cite{geniex,rxnn}. These non-idealities manifest as imprecise dot-product currents causing accuracy degradation when DNNs are mapped onto crossbars. Many previous works \cite{rxnn, geniex, vortex, liu, chen} have used frameworks to capture the impact of circuit noise or non-idealities in crossbars and proposed noise-aware retraining of DNNs to mitigate accuracy losses (see Table \ref{tab:comp}). However, \cite{rxnn, geniex, vortex, liu, chen} do not study the impact of crossbar non-idealities on the robustness of neural networks against adversarial attacks. Adversarial attacks  are structured, yet, small perturbations on the input, that fool a DNN causing high confidence misclassification. This vulnerability severely limits the deployment and potential safe-use of DNNs for real world applications~\cite{carlini, quanos}. Recently there have been investigative works such as \cite{bhattacharjee2021efficiency, roy2020robustness, roy2021intrinsic}, wherein the authors have shown that crossbar non-idealities while degrading performance, can improve the adversarial attack resilience of hardware-mapped DNNs in comparison to baseline software DNNs without additional efficiency-driven hardware optimizations (see Table \ref{tab:comp}). Furthermore, the authors in \cite{pixeld, quanos, NEURIPS2019_2ca65f58, sehwag2020hydra} show that hardware optimization techniques, such as quantization, model compression and pruning can be leveraged to improve adversarial robustness of DNNs in addition to energy-efficiency. 

\begin{table}[t]
\caption{A comparison table to show the contributions of previous works and our \textit{SwitchX} approach to mapping on non-ideal crossbars.}
\label{tab:comp}
\begin{tabular}{|ccc|ccc|}
\hline
\multicolumn{3}{|c|}{}                                                            & \multicolumn{3}{c|}{\textbf{Crossbar non-idealities included}}           \\ \hline
\multicolumn{1}{|c|}{\textbf{Work}} &
  \multicolumn{1}{c|}{\textbf{\begin{tabular}[c]{@{}c@{}}New \\ weight-mapping \\ strategy\end{tabular}}} &
  \textbf{\begin{tabular}[c]{@{}c@{}}Crossbar \\ energy-\\ efficiency\end{tabular}} &
  \multicolumn{1}{c|}{\textbf{\begin{tabular}[c]{@{}c@{}}Non-ideality \\ aware \\ performance \\ (accuracy)\end{tabular}}} &
  \multicolumn{1}{c|}{\textbf{\begin{tabular}[c]{@{}c@{}}Non-ideality \\ driven \\ adversarial \\ robustness\end{tabular}}} &
  \textbf{\begin{tabular}[c]{@{}c@{}}Noise-aware \\ training or \\ fine-tuning\end{tabular}} \\ \hline
\multicolumn{1}{|c|}{\cite{rxnn, geniex, vortex, liu, chen}}                & \multicolumn{1}{c|}{$\times$} & $\times$ & \multicolumn{1}{c|}{\checkmark}         & \multicolumn{1}{c|}{$\times$} & \checkmark          \\ \hline

\multicolumn{1}{|c|}{\cite{bhattacharjee2021efficiency, roy2020robustness, roy2021intrinsic}}                & \multicolumn{1}{c|}{$\times$} & $\times$ & \multicolumn{1}{c|}{\checkmark}         & \multicolumn{1}{c|}{\checkmark} & $\times$          \\ \hline

\multicolumn{1}{|c|}{\cite{sara, sara_new}}                & \multicolumn{1}{c|}{\checkmark}         &    \checkmark      & \multicolumn{1}{c|}{$\times$} & \multicolumn{1}{c|}{$\times$} & $\times$ \\ \hline
\multicolumn{1}{|c|}{\cite{cherupally2022improving}} & \multicolumn{1}{c|}{$\times$}         & $\times$         & \multicolumn{1}{c|}{\checkmark}         & \multicolumn{1}{c|}{\checkmark}         & \checkmark \\ \hline

\multicolumn{1}{|c|}{NEAT \cite{neat}} & \multicolumn{1}{c|}{$\times$}         & \checkmark         & \multicolumn{1}{c|}{\checkmark}         & \multicolumn{1}{c|}{\checkmark}         & \checkmark \\ \hline

\multicolumn{1}{|c|}{\textbf{Our work (\textit{SwitchX})}} & \multicolumn{1}{c|}{\textbf{\checkmark}}         & \textbf{\checkmark}         & \multicolumn{1}{c|}{\textbf{\checkmark}}         & \multicolumn{1}{c|}{\textbf{\checkmark}}         & \textbf{$\times$} \\ \hline
\end{tabular}%
\end{table}

In this work, we introduce \textit{SwitchX} based mapping of BNN weights onto ReRAM crossbars (see \figurename{~\ref{mapping}} for \textit{SwitchX} mapping), whereby we map the binarized weights in a manner such that the number of HRS synapses always dominate every crossbar instance. We show that \textit{SwitchX} mapping interfere with hardware non-idealities to yield benefits in terms of adversarial robustness when compared with normally-mapped BNNs (the baseline). This is achieved by boosting the feasibility of HRS states in crossbars to tackle crossbar non-idealities and improve the natural and adversarial performance (robustness) of BNNs for their secure deployment on edge-devices. It has been shown in earlier works that the weights of a BNN encoded as resistance states in crossbars contribute significantly to the power dissipated by the crossbars \cite{sara, Chang2015, kim2018neural}. An important by-product of \textit{SwitchX} mapping is the reduction in the overall power expended by the crossbar instances during BNN inference, owing to an increased proportion of HRS synapses.  

\textbf{Contributions:} In summary, the key contributions of this work are as follows:
\begin{enumerate}
\item We comprehensively analyse how \textit{SwitchX} mapping of binarized weights onto ReRAM crossbars manifests as increase in the robustness and adversarial stability of the mapped networks. Note, although the results in this work have been presented typically for ReRAM crossbars, the \textit{SwitchX} approach and its benefits are not limited to ReRAM crossbars and can be extended to crossbars with other NVM devices such as, PCM or FeFET. 

\item We carry out experiments on a state-of-the-art neural network architecture (VGG16)~\cite{vgg} using benchmark datasets (CIFAR-10, CIFAR-100~\cite{cifar} and TinyImagenet). We propose a novel graphical approach by plotting \textit{robustness maps} to evaluate the adversarial robustness of BNNs on  crossbars and perform a fair comparison between \textit{SwitchX} and Normal Mapping. 

\item We find that \textit{SwitchX} approach when combined with state-aware training (explained in Section~\ref{exp}) increases the feasibility of higher HRS mapping. This significantly boosts the robustness (both clean and adversarial accuracies) of the mapped BNNs. Further, \textit{SwitchX} combined with state-aware training outperforms the case when \textit{SwitchX} is combined with Adversarial training, a state-of-the-art software defense against adversarial attacks. 

\item We also carry out ablation studies and ascertain that the \textit{SwitchX} approach can unleash robustness to BNNs mapped onto crossbars with various specifications and dimensions ranging from 16$\times$16, 32$\times$32, 64$\times$64 to 128$\times$128.

\item Apart from robustness benefits, we also find that the above approach can lead to significant power savings ($\sim21-22\%$ for 16$\times$16 and 32$\times$32 crossbars using CIFAR-10 \& CIFAR-100 datasets) with respect to normally mapped BNNs on crossbars, thereby bringing in crossbar energy-efficiency. Note, as we see later (in Section \ref{switchx_power}) that the crossbar energy-efficiency via \textit{SwitchX} mapping decreases on increasing the crossbar size, we thus report the overall crossbar power savings of \textit{SwitchX} BNNs only for 16$\times$16 and 32$\times$32 crossbars.

\end{enumerate}

\section{Related Works}
\label{related}

BNNs have been widely explored as a hardware-friendly method to implement neural networks on crossbar-based architectures. Particularly, early works such as \cite{ni2017energy, chen201865nm} have proposed optimized hardware architectures using RRAM crossbars to facilitate BNN inference with higher energy-efficiency, throughput and parallelism, and have been shown to be resilient against device-level variations. Another work \cite{hirtzlin2019outstanding} has further proposed RRAM noise-aware retraining of a BNN model to further improve noise-tolerance of crossbar-mapped BNNs on error-prone crossbar platforms. Later works such as \cite{song2019rebnn} have proposed a dual-activation and dual-synapse based implementation of BNNs on RRAM crossbar-arrays (referred to as Complementary Resistive Cell (CRC) arrays). This helps perform device noise-resilient XNOR operations on the RRAM crossbars and furthers the energy-efficiency and throughput during BNN inference. Furthermore, other works focusing on energy-efficient implementations of BNN on crossbars include \cite{zheng2020lattice}, that implements an ADC-less RRAM accelerator to carry out BNN inference and outperforms standard state-of-the-art crossbar-based inference accelerators such as, ISAAC \cite{shafiee2016isaac}, Pipelayer \cite{song2017pipelayer} and FloatPIM \cite{imani2019floatpim}. However, in \cite{zheng2020lattice}, RRAM device-noise models are not included during BNN inference and hence, hardware-realistic BNN accuracies are not reported. Recent works such as \cite{tang2022hawis, nie2022cross} include mature RRAM noise models in addition to device-level variations such as stuck-at-fault defects and have proposed architectural modifications as well as noise-mitigation strategies for BNNs to achieve similar accuracy on hardware with state-of-the-art multi-precision counterparts. However, majority of the above works focus on energy-efficiency and throughput as the key goals to optimize for BNN inference on crossbar-arrays. Moreover, none of the above works have explored the area of adversarial robustness for BNNs on crossbars in presence of hardware non-idealities. 

Our work \textit{SwitchX} explores non-ideality aware adversarial robustness of BNNs in presence of resistive crossbar non-idealities in addition to device-level variations. It should be noted that although there has been a complementary work, called \textit{Non-linearity aware Training} (NEAT) \cite{neat} for non-ideality aware adversarial robustness of DNNs on crossbars, it involves additional training cost of fine-tuning a pretrained software DNN model before deploying onto non-ideal crossbars. In addition, NEAT focuses on robustness of multi-precision DNN models but does not take into account the hardware overheads of implementation of multibit weights (or conductances) with an ensemble of memristive devices (using bit-slicing) on crossbars. To this end, \textit{SwitchX} uses binary weights that are simple to program on crossbars with a single synaptic NVM device and hence, are more crossbar-friendly. Furthermore, there have been previous works such as, \cite{sara, sara_new} (see Table \ref{tab:comp}), introducing techniques that also increase the proportion of high resistance ReRAM states in crossbars leading to better crossbar energy-efficiencies. But, none of them have focused on adversarial robustness in presence of resistive crossbar non-idealities. It should also be noted that \textit{SwitchX}, unlike \cite{rxnn, geniex, vortex, liu, chen, neat, cherupally2022improving} in Table \ref{tab:comp}, is not noise-aware retraining or fine-tuning of neural networks but rather is a simple weight-mapping strategy that helps mitigate the impact of crossbar noise.

\section{Background}
\label{background}

\subsection{Memristive Crossbars and their non-idealities}
\label{memxbar}

\begin{figure}[t]
    \centering
    \includegraphics[width=0.7\linewidth]{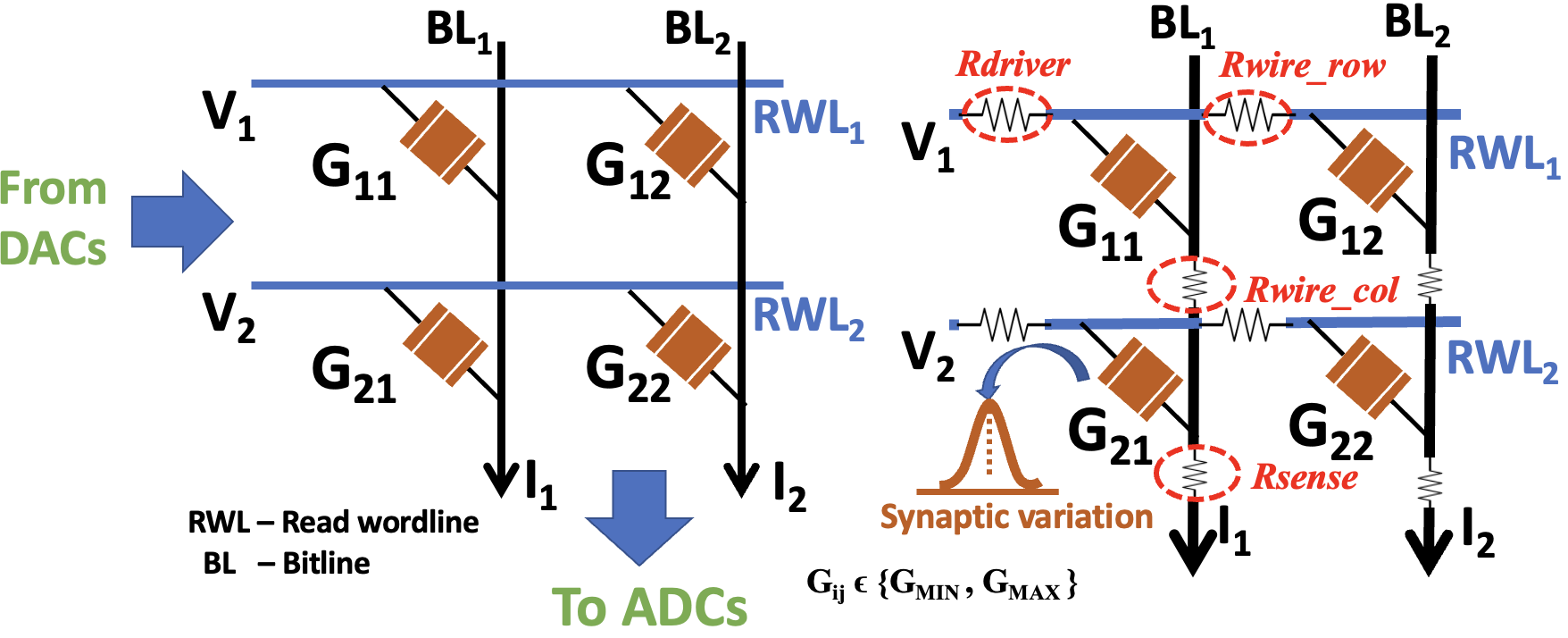}%
    \caption{(Left) A 2x2 ideal ReRAM crossbar with weights programmed as synaptic conductances ($G_{ij}$s); (Right) A 2x2 non-ideal crossbar with the resistive and device-level non-idealities marked. These non-idealities lead to imprecise dot-product currents and that manifests as accuracy degradation when neural networks are evaluated on crossbars}
    \label{xbar}
   
\end{figure}


Memristive crossbar arrays have been employed to implement \textit{Matrix-Vector-Multiplications} (MVMs) in neural networks in an analog manner. Traditionally, a crossbar (see \figurename{~\ref{xbar}} (Left)), consists of a 2D array of synaptic NVM devices interfaced with DACs, \textit{Analog-to-Digital Converters} (ADCs) and a write circuit to program the synapses. The synaptic devices at the cross-points are programmed to a particular value of conductance ($G_{ij}$). The MVM operations during inference are performed by converting the digital inputs to a neural network into analog voltages on the \textit{Read Wordlines} (RWLs) using DACs, and sensing the output current flowing through the bitlines (BLs) using the ADCs~\cite{rxnn}. For an ideal crossbar, if $V_{in}$ is the input voltage vector, $Iout_{ideal}$ is the output current vector and $G_{ideal}$ is the conductance matrix, then:

\small

\begin{equation} \label{eq:idealMVM}
Iout_{ideal}~=~V_{in}*G_{ideal}
\end{equation}
\normalsize

\textbf{Non-idealities and equivalent conductance matrix:} The analog nature of the computation leads to various non-idealities resulting in errors in the computation of MVMs. These include various linear resistive non-idealities in the crossbars. \figurename{~\ref{xbar}} (Right) shows the equivalent circuit for a crossbar array  accounting for various peripheral and parasitic non-idealities ($Rdriver$, $Rwire\_row$, $Rwire\_col$ and $Rsense$) modelled as parasitic resistances. The cumulative effect of all the non-idealities results in the deviation of the output current from its ideal value (\textit{i.e.}, equation~\ref{eq:idealMVM}), resulting in an $Iout_{non-ideal}$ vector. The relative deviation of $Iout_{non-ideal}$ from its ideal value is measured by \textit{non-ideality factor} (NF)~\cite{geniex} as:

\small

\begin{equation} \label{eq:nf}
NF~=~(Iout_{ideal}-Iout_{non-ideal})/Iout_{ideal}
\end{equation}
\normalsize

Thus, NF is a direct measure of crossbar non-idealities, \textit{i.e.} increased non-idealities  induce a greater value of NF, affecting the accuracy of the neural network mapped onto them~\cite{cxdnn, rxnn, geniex}.

In this work, we take a trained BNN model with binary weights (\{-1, +1\}) and map onto ReRAM crossbars using only two conductance states- $G_{MIN}$ (HRS) and $G_{MAX}$ (LRS) (details on mapping of negative and positive weights of the BNN along with circuit implementation have been presented in Section \ref{switchx}). From \figurename{~\ref{mapping}}, once the binarized weights of a BNN are mapped onto crossbars to obtain $G_{ideal}$, we integrate the resistive interconnect non-idealities of the crossbar and synaptic device variations to convert $G_{ideal}$ into $G_{non-ideal}$ (see Section \ref{platform}). This completes the mapping of the weights of the BNN onto a non-ideal crossbar. Thus, we have:

\small

\begin{equation} \label{eq:nonidealMVM}
Iout_{non-ideal}~=~V_{in}*G_{non-ideal}
\end{equation}
\normalsize

Unless otherwise stated, we carry out experiments on ReRAM crossbars with $R_{MIN} = 20 k\Omega$ and device ON/OFF ratio of 10~\cite{hajri2019rram}. The resistive are non-idealities as follows: $Rdriver = 1 k\Omega$, $Rwire\_row = 5 \Omega$, $Rwire\_col = 10 \Omega$ and $Rsense = 1 k\Omega$ \cite{deng2012rram, xia2016technological, rxnn, geniex}. The device-to-device variations in ReRAM conductances is modelled using a Gaussian distribution around the nominal device conductances with $\sigma/\mu = 10\%$ \cite{xia2016technological, chen2015mitigating}. For our experiments, by performing SPICE simulations with the ReRAM device model~\cite{nanoHUB.org19}, we identified that the binary analog voltages input to the ReRAM crossbars for BNN inference can be +0.1/-0.1 V to maintain linear ReRAM I-V characteristics \cite{mehonic2019simulation}. In this work, we have assumed that the correct mapping of the BNN weights to the memristive conductances (HRS or LRS) without stuck-at-fault defects is ensured post-training \cite{zhang2019handling}. We use a hardware evaluation framework in Pytorch~\cite{rxnn, bhatta} to map the BNNs onto non-ideal crossbars and investigate the cumulative impact of the resistive and device-level non-idealities on networks mapped with \textit{SwitchX} technique (explained in Section~\ref{switchx}) on the robustness of neural networks.

\subsection{Adversarial Attacks and Defense}

Neural networks are vulnerable to adversarial attacks in which the model gets fooled by applying precisely calculated small perturbations on the input~\cite{quanos}. Goodfellow \textit{et al.}~\cite{goodfellow} proposed \textit{Fast Gradient Sign Method} (FGSM) to generate adversarial attacks ($X_{adv}$) by linearization of the loss function ($L$) of the trained models with respect to the input ($X$) as shown in equation~(\ref{eq:Xadv}).

\small

\begin{equation} \label{eq:Xadv}
X_{adv}~=~X~+~\epsilon~\times~sign(\nabla_{x} L(\theta,X,y_{true}))
\end{equation}
\normalsize

Here, $y_{true}$ is the true class label for the input $X$; $\theta$ denotes
the model parameters (weights, biases etc.). The quantity $\Delta=\epsilon~\times~sign(\nabla_{x} L(\theta,X,y_{true}))$ is the net perturbation which is controlled by $\epsilon$. It is noteworthy that gradient propagation is a crucial step in unleashing an adversarial attack. Furthermore, the contribution of gradient to $\Delta$ would vary for different layers of the network depending upon the activations~\cite{quanos}. In addition to FGSM-based attacks, multi-step variants of FGSM, such as \textit{Projected Gradient Descent} (PGD)~\cite{pgd} have also been proposed that cast stronger attacks. The PGD attack, shown in equation (\ref{pgd}), is an iterative attack over $n$ steps. In each step $i$, perturbations of strength $\alpha$ are added to $X_{adv}^{i-1}$. Note, that $X_{adv}^{0}$ is created by adding random noise to the clean input $X$. Additionally, for each step, $X_{adv}^{i}$ is projected on a \textit{Norm ball} \cite{pgd}, of radius $\epsilon$. In other words, we ensure that the maximum pixel difference between the clean and adversarial inputs is $\epsilon$. 
\begin{equation}
    X_{adv} = \sum_{i=1}^{n} X_{adv}^{i-1}+\alpha ~sign(\nabla_x\mathcal{L}(\theta, X, y_{true})).
    \label{pgd}
\end{equation}

To build resilience against small adversarial perturbations, defense mechanisms such as gradient masking or obfuscation~\cite{gradmask} have been proposed. Such methods construct a model devoid of useful gradients, thereby making it difficult to create an adversarial attack. Furthermore, adversarial training \cite{goodfellow, pgd, ensat, gat} is the state-of-the-art and strongest-known software defense against adversarial attacks. Here, the training dataset is augmented with adversarial examples so that the network learns to predict them correctly. The authors in~\cite{bhattacharjee2021efficiency} have also shown that hardware non-idealities can intrinsically lead to defense via gradient obfuscation against adversarial perturbations, thereby making a neural network on hardware adversarially robust than baseline software models. 

In this work, we explore how \textit{SwitchX} mapping of BNNs on non-ideal crossbars can yield adversarial robustness when compared with Normal Mapping of BNN weights. Furthermore, we show that when \textit{SwitchX} is combined with state-aware training of BNN (discussed in Section~\ref{exp}), it intrinsically unleashes even stronger defense on hardware than adversarial training. 

\section{Methodology}
\label{sec:methodology}

\subsection{\textit{SwitchX} mapping}
\label{switchx}

For the baseline or Normal Mapping of BNNs onto crossbars, `+1' weights are mapped to $G_{MAX}$ or LRS and `-1' to $G_{MIN}$ or HRS as shown in \figurename{~\ref{mapping}} (see Normal Mapping). In this work, we perform switched-mapping of the weights of a BNN onto crossbars, termed as \textit{SwitchX}. As shown in \figurename{~\ref{mapping}} for \textit{SwitchX}, we compute the mean of the binarized weight matrix of the BNN to be mapped onto a crossbar instance. If the mean is positive-valued (mean $>$ 0), we switch the mapping with `+1' values to HRS states and `-1' values to LRS states. If mean is negative-valued or zero (mean $\leq$ 0), we do not perform any switching and conduct Normal Mapping. This approach ensures that the proportion of HRS states in a crossbar array is always higher when mapping BNN weights. As noted in \figurename{~\ref{mapping}}, the ideal conductance matrix $G_{ideal}$ obtained from the switched mapping is again converted into $G_{non-ideal}$ taking the crossbar non-idealities into account. 

\begin{figure}[t]
    \centering
    \includegraphics[width=0.5\linewidth]{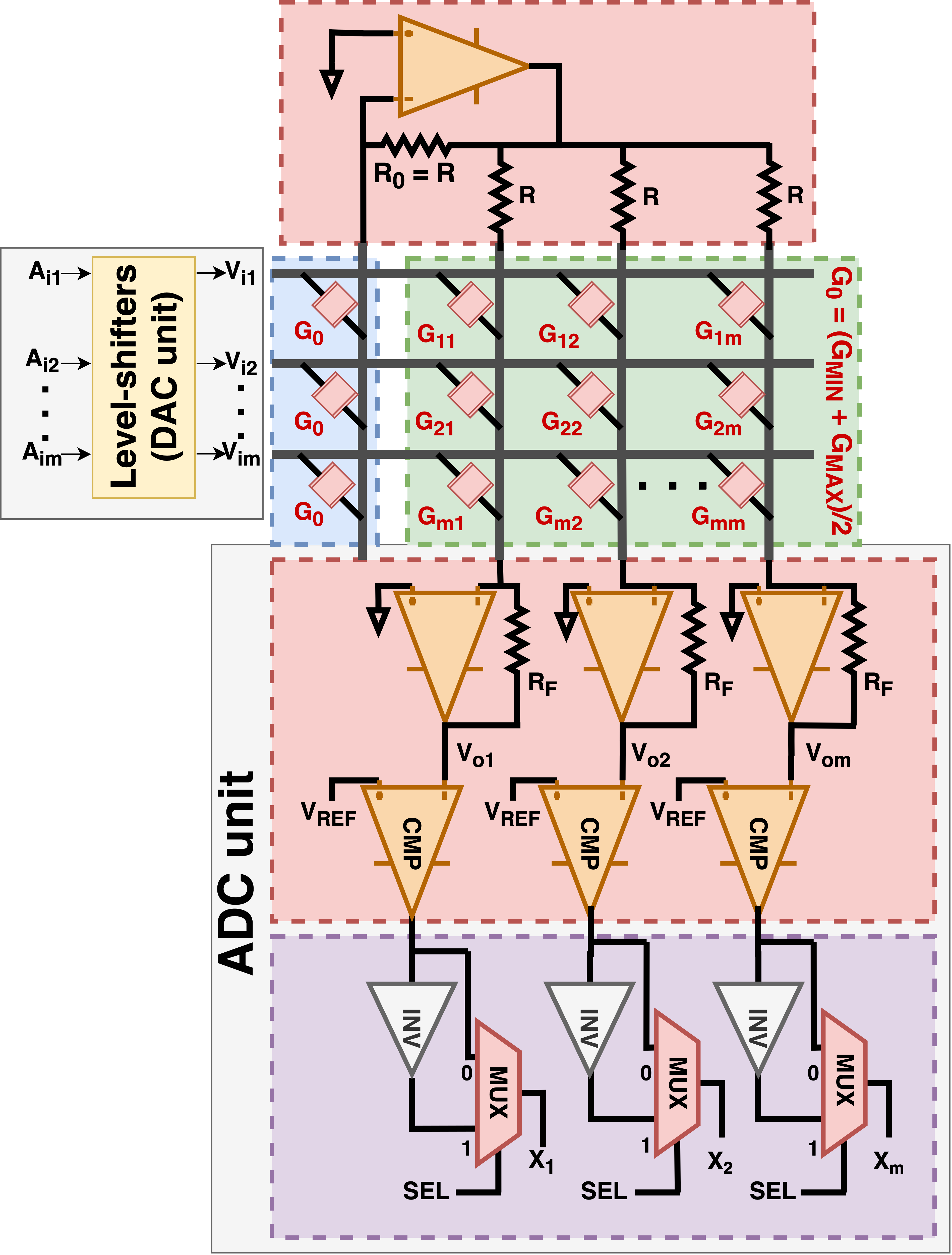}%
    \caption{Circuit for \textit{SwitchX} implementation during BNN inference shown for an m$\times$m memristive crossbar (highlighted in green). Here, $V_{i1}-V_{im}$ are analog voltages input to the crossbar and $X_1-X_m$ are the output binary activations.}
    \label{cir}
   
\end{figure}

\textbf{Circuit implementation of \textit{SwitchX} for BNN inference:} \figurename{~\ref{cir}} shows the manner in which BNN inference is carried out accurately on an m$\times$m crossbar via \textit{SwitchX} approach. Standard techniques involve a dual-crossbar approach to map neural networks onto crossbars with negative weights. However, in this work, we follow a single crossbar-based approach similar to~\cite{membin1, membin2} for mapping both positive and negative BNN weights as ReRAM conductances, thereby reducing hardware overheads. The binary digital inputs of a BNN ($A_1$-$A_m$) are converted into analog voltages ($V_{i1}$-$V_{im}$) using DACs (level-shifters producing +0.1/-0.1 V) which are input to the crossbar. The software model weights of a BNN are centered around zero, \textit{i.e.} \{-1, +1\}. In the crossbars, the binary weights of the BNN (\{-1, +1\}) are mapped as \{$G_{MIN}$, $G_{MAX}$\} synaptic conductances that are not symmetrically centered around zero. Hence, to conduct accurate BNN inference, we need to perform some transformations in hardware to ensure that the trained software weights after being mapped to conductance values still remain symmetrically centered around zero, \textit{i.e.} \{$\frac{-(G_{MAX}-G_{MIN})}{2}, \frac{+(G_{MAX}-G_{MIN})}{2}$\}. To this end, an additional column of ReRAM devices with conductances of value $G_0=\frac{(G_{MAX}+G_{MIN})}{2}$ is added (highlighted in blue) and $R_0 = R$ is set to facilitate the transformation. Note, this is viable for the case when the ReRAM devices under consideration can be programmed to a level between $G_{MIN}$ and  $G_{MAX}$ having a conductance equal to $\frac{(G_{MAX}+G_{MIN})}{2}$. In case, the ReRAM devices are bimodal, we realize $G_0=(G_{MAX}+G_{MIN})$ using a parallel combination of two devices- one in HRS state and the other in LRS state, and set $R_0 = \frac{R}{2}$. The resulting crossbar currents are sensed by the ADC unit consisting of transconductance amplifiers (generating voltages $V_{o1}$-$V_{om}$) followed by comparators (with $V_{REF}=0$) to produce binarized activations. However, for weights mapped onto crossbars via \textit{SwitchX} transformation, wherein HRS and LRS states are interchanged for the case when mean $>$ 0, a corresponding inverse transformation is also necessary in the end to ensure the correct activations ($X_1$-$X_m$) at the output. This is done by the digital module consisting of inverters and multiplexers (highlighted in violet). The $SEL$ signal input to the multiplexers is set to `0' for mean $>$ 0 scenario to conduct switched-mapping and `1' for mean $\leq$ 0 scenario to conduct Normal Mapping to produce the correct output activations ($X_1$-$X_m$). Effectively, during \textit{SwitchX} implementation, based on the value of mean, the analytical operations are governed by equations (\ref{eq:mac-normal}) \& (\ref{eq:mac-sa}). 

For mean $\leq$ 0,
\small

\begin{equation} \label{eq:mac-normal}
\sum\limits_{i}^{}a_i*w_{ij}~=~\sum\limits_{i}^{}V_i*G_{ij} - \sum\limits_{i}^{}V_i*G_0
\end{equation}
\normalsize

Here, the subtraction operation is needed to ensure that the trained binarized weights after being mapped to conductance values in the crossbars remain symmetrically centered around zero, \textit{i.e.} \{$\frac{-(G_{MAX}-G_{MIN})}{2}, \frac{+(G_{MAX}-G_{MIN})}{2}$\}. 
Note, the above equation is also true for normally mapping a BNN irrespective of the value of mean~\cite{membin1}. 

For mean $>$ 0,
\small

\begin{equation} \label{eq:mac-sa}
\sum\limits_{i}^{}a_i*w_{ij}~=~-[\sum\limits_{i}^{}V_i*G_{ij} - \sum\limits_{i}^{}V_i*G_0]
\end{equation}
\normalsize

Basically, for switched-mapping in equation \ref{eq:mac-sa}, the additional inverse operation with respect to equation \ref{eq:mac-sa} is emulated using the digital inverse transformation module (highlighted in violet in Fig. \ref{cir}).

\begin{figure}[t]
    \centering
    \includegraphics[width=0.3\linewidth]{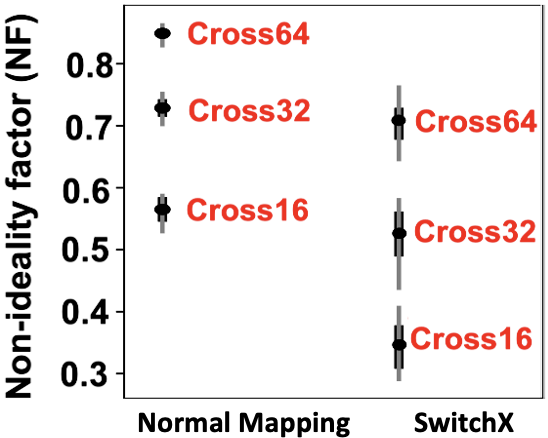}%
    \caption{Box Plot of error bars to show variations in NF for various crossbar sizes for Normal and \textit{SwitchX} mapping}
    \label{nf}
   
\end{figure}

\textbf{Trends for non-ideality factor (NF): }\figurename{~\ref{nf}} shows the variation in NF for different crossbar sizes for a given binary weight matrix with higher proportion of `+1' values (mean $>$ 0) for a fully-connected layer of a neural network and a given set of input voltages drawn randomly from a uniform distribution. It is observed that NF increases with increasing crossbar size (16$\times$16 to 64$\times$64). Further for a given crossbar size, \textit{SwitchX} results in a decrease in the value of NF with respect to the case of Normal Mapping. This occurs because \textit{SwitchX} increases the effective resistance of a crossbar array, thereby minimizing the effect of interconnect parasitic resistances and device-level variations, by increasing the proportion of HRS states. \textit{Thus, BNNs mapped onto crossbars via \textit{SwitchX} approach would suffer less interference from non-idealities on MVM operations which would thereby lead to lesser accuracy degradation. }

\subsection{Hardware evaluation framework}
\label{platform}

\begin{figure*}[t]
    \centering
    \includegraphics[width=\linewidth]{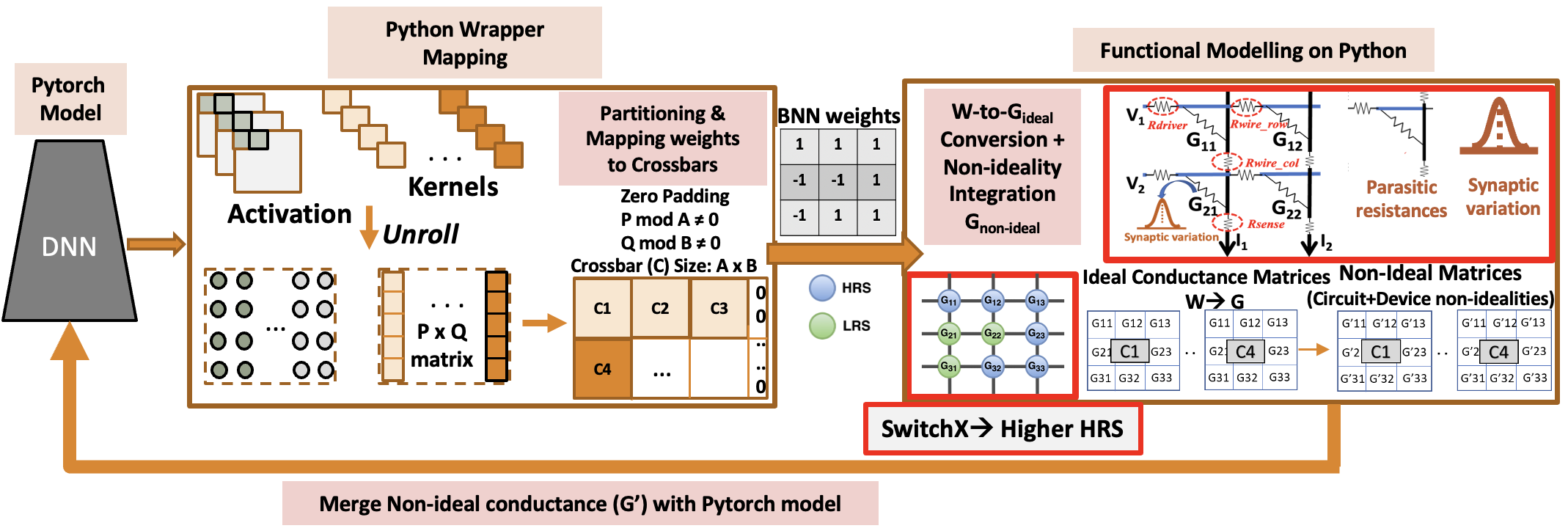}%
    \caption{Hardware evaluation platform for mapping the BNN parameters onto non-ideal crossbars following \textit{SwitchX} technique}
    \label{main-switchx}
    
\end{figure*}

As discussed in Section~\ref{memxbar}, we use a framework in Pytorch, based on \textit{RxNN} \cite{rxnn}, to map trained BNNs onto non-ideal memristive crossbars via \textit{SwitchX} technique and investigate the cumulative impact of the circuit and device-level non-idealities on their adversarial robustness. \figurename{~\ref{main-switchx}} illustrates the overall simulation framework that is used for non-ideal crossbar evaluation using \textit{SwitchX} technique. The entire platform, being based on Python, enables better integration between the software model and the simulation framework. In the platform, a Python wrapper is built that unrolls each and every convolution operation in the software BNN into MAC operations. The matrices obtained are then zero-padded (in case the size of the weight matrix is not an exact multiples of the crossbar size) and partitioned into multiple crossbar instances consisting of the binarized weights. The next stage (functional modelling) of the platform converts the binarized weights $W$ to suitable conductance states $G$ (HRS or LRS as marked in green and blue colours respectively) using \textit{SwitchX} approach based on proportions of `+1's and `-1's in the crossbar instances. Thereafter, the circuit-level parasitic non-idealities are integrated via circuit laws (\textit{Kirchoff's laws}) and linear algebraic operations written in Python \cite{bhatta}. This integration is similar to the numerical operations adopted in the \textit{RxNN} framework, which is a recent work that accurately models circuit non-idealities while evaluating DNNs on crossbars during inference. \textit{RxNN} has been shown to closely match the results obtained using SPICE models for interconnect parasitics and achieves significantly high speed up during DNN inference on non-ideal crossbars. Further, the ReRAM device variations are included with gaussian profiling. The default specifications of the NVM devices as well as the values of non-idealities have been listed in Table \ref{def_table}, which are used for our experiments. The non-ideal synaptic conductances $G'$ are then integrated into the original Pytorch based BNN model to conduct inference. This framework, thus, enables us to analyze the impact of intrinsic crossbar non-idealities on mapping BNNs with \textit{SwitchX}.  

\begin{table}[t]
\centering
\caption{Table showing default crossbar and NVM device specifications for non-ideality aware performance evaluation of BNNs}
\label{def_table}
\begin{tabular}{|ll|ll|}
\hline
\multicolumn{2}{|c|}{\textbf{Synapse characteristics}} & \multicolumn{2}{c|}{\textbf{Non-idealities}}  \\ \hline
\multicolumn{1}{|l|}{\textbf{Parameter}} & \textbf{Value} & \multicolumn{1}{l|}{\textbf{Parameter}}   & \textbf{Value}          \\ \hline
\multicolumn{1}{|l|}{$R_{MIN}$}            & 20 k$\Omega$   & \multicolumn{1}{l|}{$Rdriver$, $Rsense$}      & 1 k$\Omega$, 1 k$\Omega$  \\ \hline
\multicolumn{1}{|l|}{$R_{MAX}$}            & 200 k$\Omega$  & \multicolumn{1}{l|}{$Rwire\_row$, $Rwire\_col$} & 5 $\Omega$, 10 $\Omega$ \\ \hline
\multicolumn{1}{|l|}{ON/OFF ratio}         & 10        & \multicolumn{1}{l|}{Synaptic variation} & 10\% \\ \hline
\end{tabular}
\end{table}

\section{Experiments}
\label{exp}

We conduct experiments on a VGG16 BNN architecture with benchmark datasets- CIFAR-10, CIFAR-100 and TinyImagenet. The CIFAR-10 and CIFAR-100 datasets consist of 50,000 training and 10,000 test RGB images of size 32$\times$32 belonging to 10 and 100 classes, respectively. The TinyImagenet dataset is a larger and complex dataset consisting of RGB images of size 64$\times$64. It is a subset of the Imagenet dataset having 100,000 training images and 10,000 test images from 200 different classes. After training the BNNs on software, we launch FGSM or PGD attacks by adding adversarial perturbations to the clean test inputs and record the adversarial accuracies in each case. This forms our baseline software models (first two rows in \tablename{~\ref{baseline}}).  

\textbf{State-aware training: }In standard scenarios, the binarization of weights during software-training occurs with respect to a threshold ($\delta$) value of 0.0. That is, a weight value greater (or lesser) than 0.0 is quantized as `+1' (or `-1'), respectively. State-aware training is a method to make the distribution of binarized weights across a channel more non-uniform~\cite{sara}. In this approach, the threshold ($\delta$) is a hyperparameter assuming a positive value when there are more negative weights in a channel and a negative value otherwise. The weights are now quantized according to the new threshold value. For the VGG16 BNN, we found that a value $|\delta| = 1e-3$ for CIFAR-10 \& TinyImagenet datasets and $|\delta| = 5e-4$ for CIFAR-100 dataset were optimal to retain the performance of the BNNs upon state-aware training on software and hence, selected for the upcoming experiments. Subsequently, \textit{SwitchX}-mapping on such networks increases the proportion of HRS states in the crossbar instances (implying greater reduction in crossbar non-ideality factor). State-aware training on VGG16 BNN with CIFAR-10 dataset yields a baseline software model as shown in \tablename{~\ref{baseline}} (fourth row). Data for VGG16/CIFAR-100 and VGG16/TinyImagenet with state-aware training have not been shown for brevity. We extensively analyse the benefits of this approach in the upcoming sections. 

\textbf{Modes of adversarial attack: }For the adversarial attacks (FGSM/PGD) on the crossbar-mapped models of the BNNs, we consider two modes:
\begin{enumerate}
\item \textit{Software-inputs-on-hardware (SH) mode}: The adversarial perturbations for each attack are created using the software-based baseline's loss function and then added to the clean input that yields the adversarial input. The generated adversaries are then fed to the crossbar-mapped BNN. This is a kind of black-box adversarial attack.

\item \textit{Hardware-inputs-on-hardware (HH) mode}: The adversarial inputs are generated for each attack using the loss from the crossbar-based hardware models. This is a kind of white-box adversarial attack. It is evident that HH perturbations will incorporate the intrinsic hardware non-idealities and thus will cast stronger attacks than SH.
\end{enumerate}

For all FGSM attacks in this work, we have $\epsilon =\{0.05, 0.1, 0.15, 0.2, 0.25, 0.3\}$. Also, we consider PGD attacks iterated over 7 steps with $\alpha = 2/255$ and $\epsilon =\{2/255, 4/255, 8/255, 16/255, 32/255\}$. Note, for all upcoming analyses which include the impact of interconnect parasitic non-idealities, to have a feasible runtime for the simulations during BNN inference that are out by first mapping weights on crossbars via \textit{SwitchX} method and then integrating with non-idealities and variations (see \figurename{~\ref{main-switchx}}), we have shown results on crossbar sizes of 16$\times$16 or 32$\times$32.

\section{Results and Discussion}
\label{result}

\begin{table*}[t]
\caption{Baseline software models using VGG16 BNN with CIFAR-10, CIFAR-100 and TinyImagenet datasets subjected to FGSM attack}
\label{baseline}
\resizebox{\linewidth}{!}{%
\begin{tabular}{|c|c|c|c|c|c|c|c|}
\hline
\textbf{Dataset} & \textbf{Clean Accuracy (\%)} & \multicolumn{6}{c|}{\textbf{Adversarial accuracies (\%)}} \\ \cline{3-8} 
 &  & \textbf{$\epsilon$ = 0.05} & \textbf{$\epsilon$ = 0.1} & \textbf{$\epsilon$ = 0.15} & \textbf{$\epsilon$ = 0.2} & \textbf{$\epsilon$ = 0.25} & \textbf{$\epsilon$ = 0.3} \\ \hline
\textbf{CIFAR-10} & 88.92 & 44.63 & 39.81 & 35.87 & 32.7 & 30.58 & 28.17 \\ \hline
\textbf{CIFAR-100} & 55.04 
& 15.97 & 13.29 & 11.51 & 10.01 & 8.82 & 7.98 \\ \hline
\textbf{TinyImagenet} & 47.4 & 6.28 & 5.47 & 4.8 & 4.25 & 3.93 & 3.7 \\ \hline
\textbf{CIFAR-10 with state-aware training ($|\delta| = 1e-3$)} & 87.55 & 46.61 & 40.63 & 35.64 & 31.82 & 28.86 & 26.38 \\ \hline
\end{tabular}%
}
\end{table*}

\subsection{Reduction in Adversarial Noise Sensitivity}

Earlier works~\cite{quanos, defquant} have identified a metric termed as \textit{Adversarial Noise Sensitivity} (ANS) to quantify the sensitivity of each layer of a neural network to adversarial perturbations. ANS for a layer $l$ in a neural network, subjected to an adversarial attack, is defined in terms of an error ratio as follows:

\small

\begin{equation} \label{eq:ans}
ANS_{l}~=~\frac{||A_{adv}^l-A^l||_2}{||A^l||_2} 
\end{equation}
\normalsize

where, $A^l$ and $A_{adv}^l$ are respectively the clean and adversarial activation values of the layer $l$. ANS is a simple metric to evaluate how each layer contributes to the net adversarial perturbation
during the gradient propagation. 

\begin{figure}[t]
    \centering
    \includegraphics[width=0.4\linewidth]{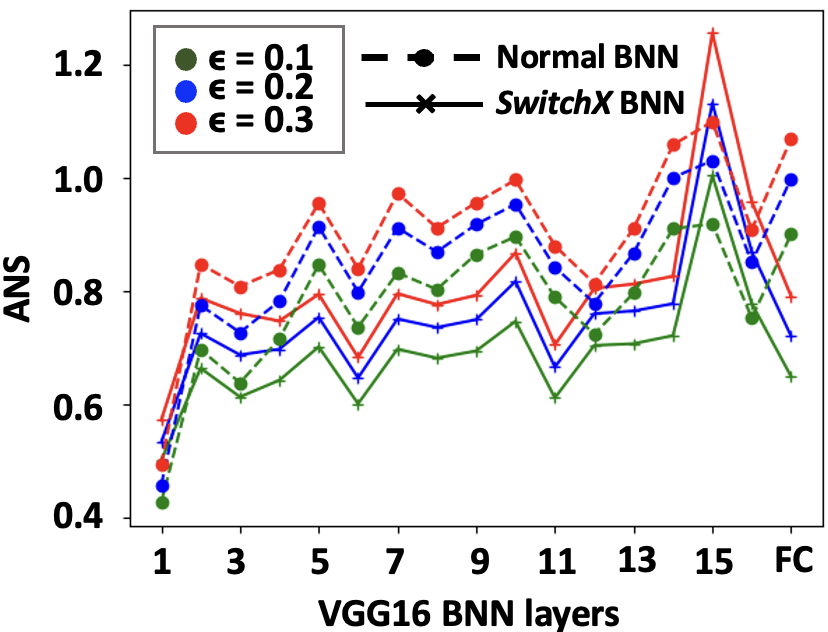}%
    \caption{ANS values plotted across different layers of VGG16 BNN trained on CIFAR-10 for Normal Mapping and \textit{SwitchX} mapping on 32$\times$32 crossbars. The \textit{SwitchX} BNN is trained via state-aware training. Here, layers 3, 6, 11 and 16 denote the Pooling layers and FC marks the final fully-connected layer. Results are shown for HH mode of FGSM attack with $\epsilon~=~0.1,0.2,0.3$.}
    \label{ans}
   
\end{figure}

In \figurename{~\ref{ans}} we show results for a VGG16 BNN (trained using state-aware training) mapped via \textit{SwitchX}, and a standard VGG16 BNN mapped normally onto 32$\times$32 crossbars ($R_{MIN} = 20 k\Omega$ and ON/OFF ratio = 10), both adversarially perturbed using HH mode of FGSM attack. We plot the ANS values for all the convolutional layers and the final fully-connected layer (marked as FC). For the case of \textit{SwitchX}-mapped BNN we observe lower ANS values implying reduced error amplification effect~\cite{defquant} and hence, lesser sensitivity and better stability to the induced adversarial perturbations. However, the authors in~\cite{quanos} also agree that a lower value of ANS across the layers of a neural network would not necessarily imply improved adversarial robustness under all circumstances. Hence, in Section~\ref{robustness}, we propose a novel graphical approach by plotting \textit{`robustness maps'} to evaluate  the  adversarial robustness of BNNs upon \textit{SwitchX} mapping onto non-ideal crossbars. 

\subsection{Robustness analysis}
\label{robustness}
\begin{figure*}[t]
    \centering
    \includegraphics[width=\linewidth]{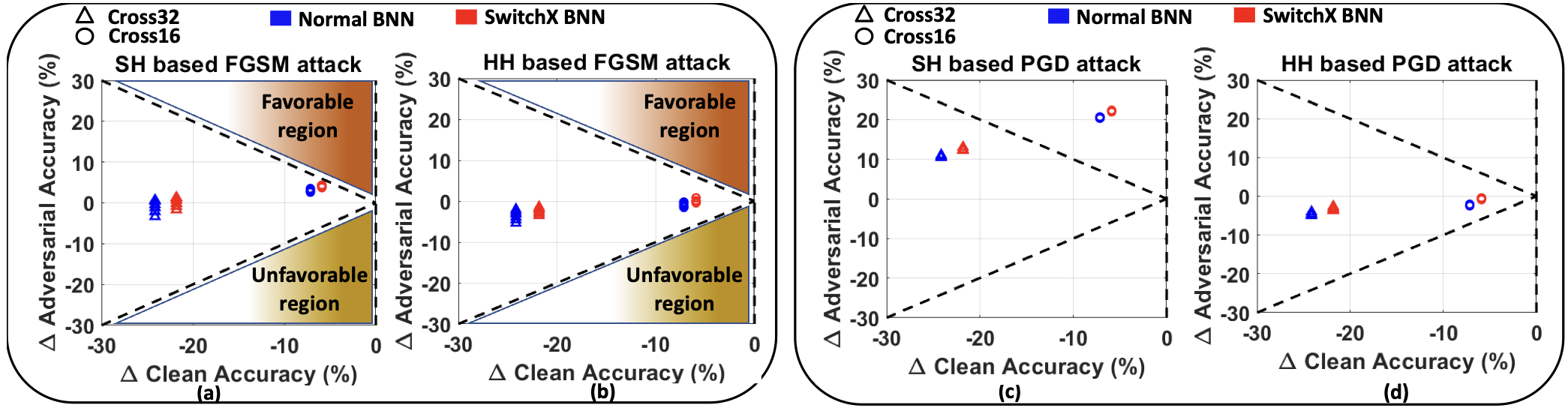}%
    \caption{(a)-(b) Robustness maps for VGG16 BNN using CIFAR-100 dataset for SH and HH modes of FGSM attack respectively; (c)-(d) Robustness maps for VGG16 based BNN using CIFAR-100 dataset for SH and HH modes of PGD attack respectively}
    \label{cifar100}
   
\end{figure*}

\begin{figure*}[t]
    \centering
    \includegraphics[width=\linewidth]{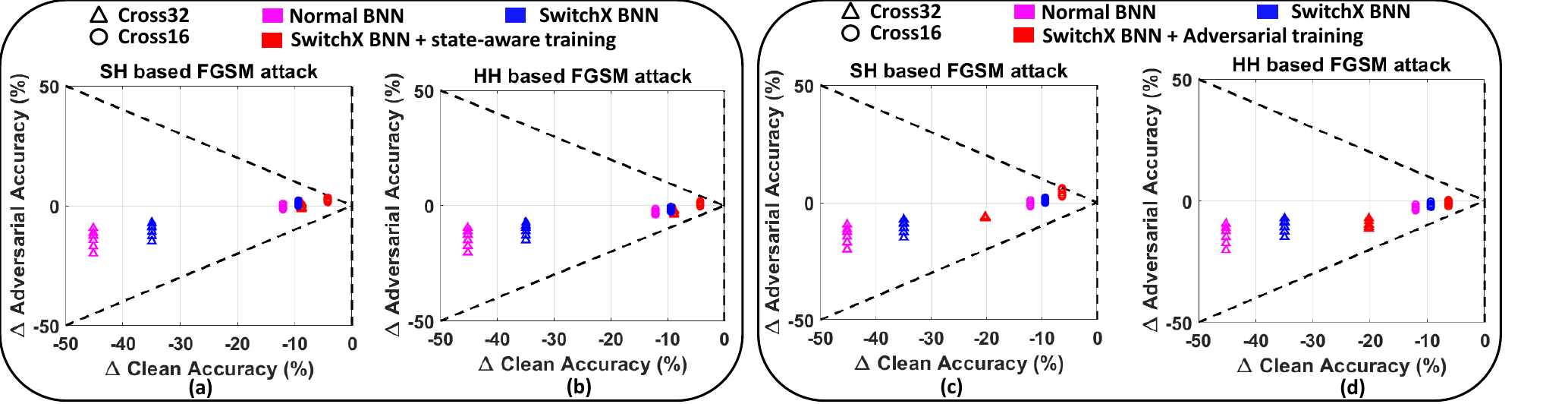}%
    \caption{(a)-(b) Robustness maps for VGG16 based BNN using CIFAR-10 dataset for SH and HH modes of FGSM attack respectively considering the case of \textit{SwitchX} combined with state-aware training; (c)-(d) Robustness maps for VGG16 based BNN using CIFAR-10 dataset for SH and HH modes of FGSM attack respectively considering the case of \textit{SwitchX} combined with adversarial training}
    \label{cifar10}
   
\end{figure*}

Earlier approaches to quantify adversarial robustness of neural networks have been based on \textit{Adversarial Loss} (AL) metric which is the difference between the clean accuracy and the adversarial accuracy for a given value of $\epsilon$~\cite{bhatta,quanos,pixeld}. Note, clean accuracy is the natural accuracy of a neural network when not under attack. A reduction in the value of AL is said to improve the adversarial robustness of the network. However, AL is not always a suitable metric for the assessment of robustness since reduction in AL does not convey whether the cause is a decrease in clean accuracy or an increase in adversarial accuracy or both. For instance, while evaluating a network with CIFAR-10 dataset, we find the clean and the adversarial accuracies (for a particular $\epsilon$) to be $10\%$ each. Then the value of AL is zero implying that the network is exceptionally robust. However, this is absurd since a $10\%$ accuracy is random for CIFAR-10 dataset implying that the network is arbitrarily predicting and is not trained.  

In this work, we evaluate robustness of the mapped BNNs on crossbars graphically as shown in \figurename{~\ref{cifar100}} and \figurename{~\ref{cifar10}}. For a specific mode of attack (SH or HH) and a given crossbar size, we plot $\Delta~Clean~Accuracy$, the difference between clean accuracy of the mapped network in question and the corresponding clean accuracy of the software baseline, on the x-axis. $\Delta~Adversarial~Accuracy$ (for a particular $\epsilon$ value) which is the difference between the adversarial accuracy of the mapped network in question and the corresponding adversarial accuracy of the software baseline is plotted on the y-axis. We term this plot as a \textit{`robustness map'}. The value of $\Delta~Clean~Accuracy$ is negative since BNNs when mapped on hardware suffer accuracy loss owing to non-idealities. The region bounded by the line $y=-x$ and the y-axis denotes that the absolute increase in the adversarial accuracy is higher than the absolute degradation in the clean accuracy. If a point lies in this region closer to the y-axis, it implies greater adversarial accuracy with lower clean accuracy loss and hence, greater robustness. Therefore, this is our \textit{favorable region}. As we move farther away from the y-axis in the favorable region, the robustness reduces as the neural network suffers from high loss of clean accuracy, and this has been shown by the variation in color gradient from dark to light-brown (dark shade implying higher robustness). 
Likewise, the region bounded by the line $y=x$ and the y-axis is where the mapped-network is highly vulnerable to adversarial attacks and hence, the \textit{unfavorable region}. Here, as we move farther away from the y-axis in the unfavorable region, the robustness reduces and the degree of \textit{unfavorability} increases. This has been shown by the variation in color gradient from dark to light-yellow (darker shade implying higher robustness). 

This approach to assess robustness of a network is comprehensive and accurate since, it takes into account the cumulative impact of both clean accuracy and adversarial accuracy (which is a strong function of the clean accuracy). The closer a point is to the region marked with dark-brown colour, the better the robustness of the neural network. Note, in \figurename{~\ref{cifar100}} and \figurename{~\ref{cifar10}}, the circular points correspond to mappings on 16$\times$16 crossbars while the triangular points correspond to mappings on 32$\times$32 crossbars. All the details pertaining to the baseline software models have been listed in \tablename{~\ref{baseline}}. Furthermore, the results in \figurename{~\ref{cifar100}} and \figurename{~\ref{cifar10}} are for BNNs mapped onto crossbars having ReRAM device ON/OFF ratio of 10 with $R_{MIN} = 20k\Omega$. 

\figurename{~\ref{cifar100}} shows the robustness maps for BNNs based on VGG16 network with CIFAR-100 dataset for both SH and HH modes of attack. \figurename{~\ref{cifar100}}(a) and \figurename{~\ref{cifar100}}(b) pertain to FGSM attack with $\epsilon$ varying from 0.05 to 0.3 with step size of 0.05. We find that \textit{SwitchX} imparts greater clean accuracy ($\sim3\%$ for 32$\times$32 crossbar) as well as better adversarial accuracies on hardware for both modes of attack, with the points corresponding to \textit{SwitchX} situated closer to the dark-brown portion favorable region than the corresponding points for Normal Mapping. This is a consequence of the reduction in non-ideality factor in case of \textit{SwitchX} with respect to Normal Mapping as discussed in Section~\ref{memxbar}. Note, the points for 32$\times$32 crossbars are situated farther from the favorable region than the corresponding points for 16$\times$16 crossbars. This is owing to greater non-idealities that exist in case of a 32$\times$32 crossbar than a 16$\times$16 crossbar (\figurename{~\ref{nf}}). We further observe that the points for \textit{SwitchX BNN} for a given crossbar size are more closely packed than the corresponding points for Normal BNN. This implies that even on increasing the perturbation strength ($\epsilon$), lesser adversarial loss is observed in case of \textit{SwitchX BNN}.

\figurename{~\ref{cifar100}}(c) and \figurename{~\ref{cifar100}}(d) also present similar results but for a PGD attack with $\epsilon$ varying from 2/255 to 32/255 with step size of 2/255. In this case, the robustness is very high for SH mode of attack as compared to HH mode of attack, with points corresponding to 16$\times$16 crossbars situated inside the favorable region. Similar to the case of FGSM attack, \textit{SwitchX} outperforms a Normal BNN in terms of robustness for both modes of attack. Here, points for different $\epsilon$ values, given a style of mapping and crossbar size, are more closely packed than the corresponding points of FGSM attack. This implies that hardware non-idealities interfere more with PGD attacks than FGSM attacks resulting in lesser accuracy degradation.

\begin{figure}[t]
    \centering
    \subfloat[]{
    \includegraphics[width=.45\linewidth]{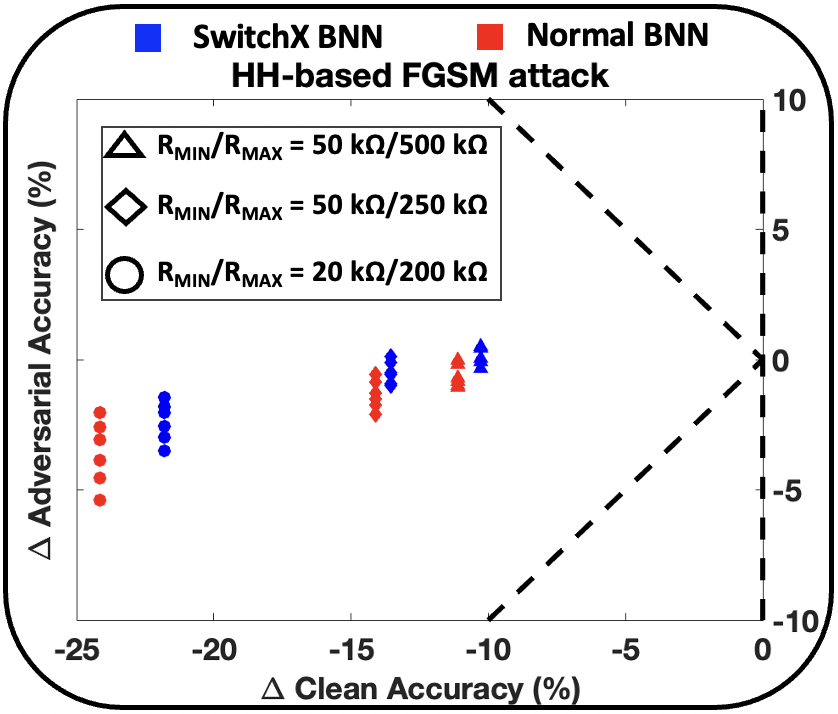}%
    }
    \subfloat[]{
    \includegraphics[width=.45\linewidth]{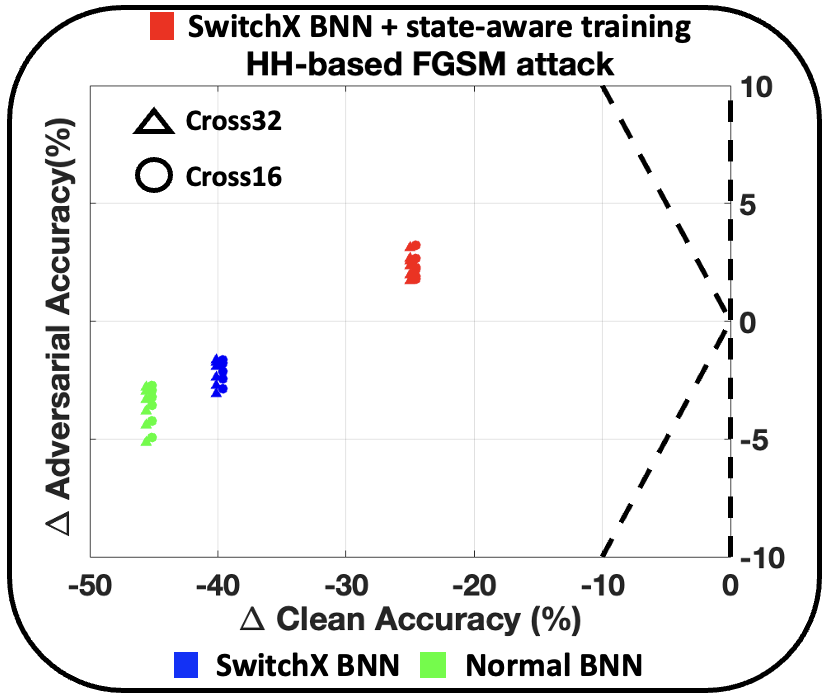}
    }
    \caption{Robustness maps for VGG16 BNN with- (a) CIFAR-100 dataset for HH mode of FGSM attack mapped onto 32$\times$32 crossbars with $R_{MIN}/R_{MAX}= 20 k\Omega/200k\Omega$ (circle), $R_{MIN}/R_{MAX}= 50 k\Omega/250k\Omega$ (diamond) and $R_{MIN}/R_{MAX}= 50 k\Omega/500k\Omega$ (triangle); (b) TinyImagenet dataset for HH mode of FGSM attack mapped onto 32$\times$32 crossbars (triangle) and 16$\times$16 crossbars (circle) with $R_{MIN}/R_{MAX}= 20 k\Omega/200k\Omega$}
    \label{rmin}
\end{figure}

\textbf{Effect of varying $R_{MIN}$ of ReRAM devices:} Previous works such as \cite{geniex, bhattacharjee2021efficiency} have shown that the impact of resistive crossbar non-idealities (or NF) decreases upon increasing $R_{MIN}$ of memristive devices. To this end, we plot robustness maps in Fig. \ref{rmin}(a) to compare Normal and \textit{SwitchX} BNNs upon increasing $R_{MIN}$ from 20 $k\Omega$ to 50 $k\Omega$. We find that for both Normal and \textit{SwitchX} BNNs, data points corresponding to $R_{MIN} = 50 k\Omega$ (triangle) are closer to the favorable region than for $R_{MIN} = 20 k\Omega$ (circle) for the same ON/OFF ratio of 10, signifying reduced impact of non-idealities on the crossbar-mapped network models. Further, for $R_{MIN} = 50 k\Omega$, \textit{SwitchX} BNNs show performance improvement in terms of robustness over Normal BNNs ($\sim1\%$ improvement in clean accuracy and adversarial accuracies for HH based FGSM attack on 32$\times$32). Also, in Fig. \ref{rmin}(a), we increase $R_{MIN}$ from 20 $k\Omega$ to 50 $k\Omega$ and reduce the device ON/OFF ratio to 5 (diamond-shaped points). We find that even at a lower ON/OFF ratio of 5, \textit{SwitchX} BNN outperforms the corresponding Normal BNN in terms of robustness. Another important observation from Fig. \ref{rmin}(a) is that robustness of a neural network on non-ideal crossbars is a stronger function of $R_{MIN}$ than $R_{MAX}$. This is because for both \textit{SwitchX} and Normal BNNs, improvement in robustness is greater on traversing from the circular to the diamond-shaped points than from the diamond-shaped points to the triangular points. 

\textbf{Efficacy of \textit{SwitchX} combined with state-aware training:} \figurename{~\ref{cifar10}} shows the robustness maps for BNNs based on VGG16 network with CIFAR-10 dataset for both SH and HH modes of FGSM attack. Here, we find that there is a significant benefit in terms of improvement in clean accuracy ($\sim10\%$) and adversarial accuracies ($\sim2-5\%$) due to \textit{SwitchX} for a 32$\times$32 crossbar with respect to Normal Mapping. Similar to the case for CIFAR-100 dataset, we find that \textit{SwitchX} outperforms a Normal BNN in terms of robustness for both modes of attack. We now analyse cases when \textit{SwitchX} is combined with state-aware training as well as adversarial training. The results for FGSM attack are summarized:

\subsubsection{With Adversarial Training}

From \figurename{~\ref{cifar10}}(c) and \figurename{~\ref{cifar10}}(d), we find that \textit{SwitchX} combined with adversarial training significantly boosts the robustness of the mapped BNN both in terms of clean as well as adversarial accuracy improvement. For 16$\times$16 crossbar, the points (in red) for different $\epsilon$ lie in the close vicinity of the boundary of the favorable region while for 32$\times$32 crossbar, the rise in clean accuracy with respect to Normal BNN is very high ($\sim20\%$). Furthermore, given a style of mapping and crossbar size, the points for adversarial training are more closely packed than the corresponding points of standalone \textit{SwitchX BNN} or Normal BNN, implying lesser accuracy losses on increasing perturbation strength of attack. 

\subsubsection{With State-aware Training}

As discussed in Section~\ref{exp}, state-aware mapping combined with \textit{SwitchX} can lead to increase in the proportion of HRS states in the crossbar instances, and thus a reduction in the non-ideality factor of the crossbars. From \figurename{~\ref{cifar10}}(a) and \figurename{~\ref{cifar10}}(b), we find that this approach significantly boosts the robustness of the mapped BNN both in terms of clean as well as adversarial accuracy improvements. For both 16$\times$16 and 32$\times$32 crossbars, the points (in red) for different $\epsilon$ lie in the close vicinity of the boundary of the favorable region. We find that for 32$\times$32 crossbar, the rise in clean and adversarial accuracies is so large that it becomes comparable to the case of standalone \textit{SwitchX BNN} mapped on a smaller 16$\times$16 crossbar. Overall, we find the rise in clean and adversarial accuracies is $\sim35\%$ and $\sim6-16\%$ greater than Normal BNN, respectively. In this case also, the points for different $\epsilon$ values, given a style of mapping and crossbar size, are more closely packed than the corresponding points of standalone \textit{SwitchX BNN} or Normal BNN, implying lesser accuracy losses on increasing perturbation strength of attack. Interestingly, we find that this approach emerges as a stronger defense against adversarial attacks than \textit{SwitchX} combined with adversarial training (a state-of-the-art software defense), the defensive action being more pronounced for larger crossbar sizes (32$\times$32 in \figurename{~\ref{cifar10}}(a) and \figurename{~\ref{cifar10}}(b)). 

Furthermore, we observe similar results in \figurename{~\ref{rmin}}(b) using the large and complex TinyImagenet dataset with VGG16 BNN. For crossbar sizes of 16$\times$16 and 32$\times$32, we find \textit{SwitchX} BNNs to outperform the corresponding Normal BNNs in terms of both clean and adversarial accuracies. There is $\sim5.5\%$ improvement in clean accuracies and $\sim1-2\%$ improvement in FGSM adversarial accuracies for HH mode of attack. With state-aware training and \textit{SwitchX} mapping combined, the improvements on 16$\times$16 and 32$\times$32 crossbars shoot upto $\sim21.2\%$ for clean accuracies and $\sim4.5-8.3\%$ for adversarial accuracies. 

\begin{figure}[t]
    \centering
    \includegraphics[width=0.7\linewidth]{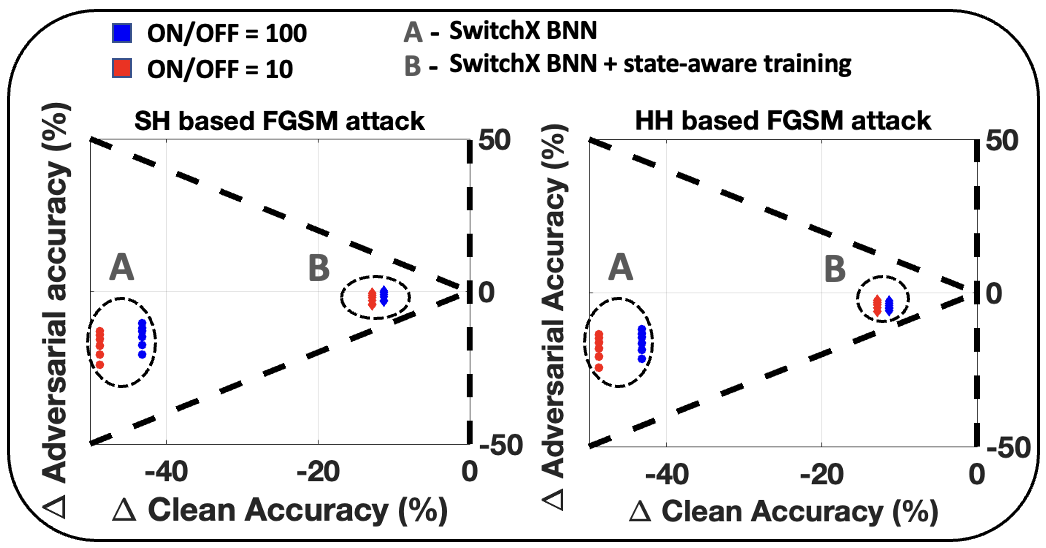}%
    \caption{Robustness maps for VGG16 BNN with CIFAR-10 dataset for SH (Left) and HH (Right) modes of FGSM attack respectively mapped onto 32$\times$32 crossbars with device ON/OFF ratios of 10 (red) and 100 (blue) for both standalone \textit{SwitchX} BNN and \textit{SwitchX} BNN + state-aware training}
    \label{on-off}
    
\end{figure}

\textbf{Effect of varying device ON/OFF ratio in crossbars:} The results in~\cite{geniex} show that on increasing the ReRAM device ON/OFF ratio (increasing the value of HRS at a constant value of LRS) in crossbars, the non-ideality factor decreases. This should translate to robustness benefits for our \textit{SwitchX} approach that in itself increases the feasibility of HRS states in crossbars. In \figurename{~\ref{on-off}}, we map BNNs onto 32$\times$32 crossbars having device ON/OFF ratios of 10 and 100 respectively. We then compare the robustness of a standalone \textit{SwitchX} BNN and a \textit{SwitchX} BNN combined with state-aware training for a VGG16 network with CIFAR-10 dataset for both cases. We find that mapping BNNs on crossbars having higher ON/OFF ratios boosts the robustness of the mapped BNNs against both SH and HH based adversarial attacks ($\sim2-4\%$ for standalone \textit{SwitchX}). Further, there are $\sim6\%$ and $\sim2\%$ improvements in clean accuracy respectively for standalone \textit{SwitchX} and \textit{SwitchX} combined with state-aware training.

\begin{figure}[t]
    \centering
    \includegraphics[width=0.75\linewidth]{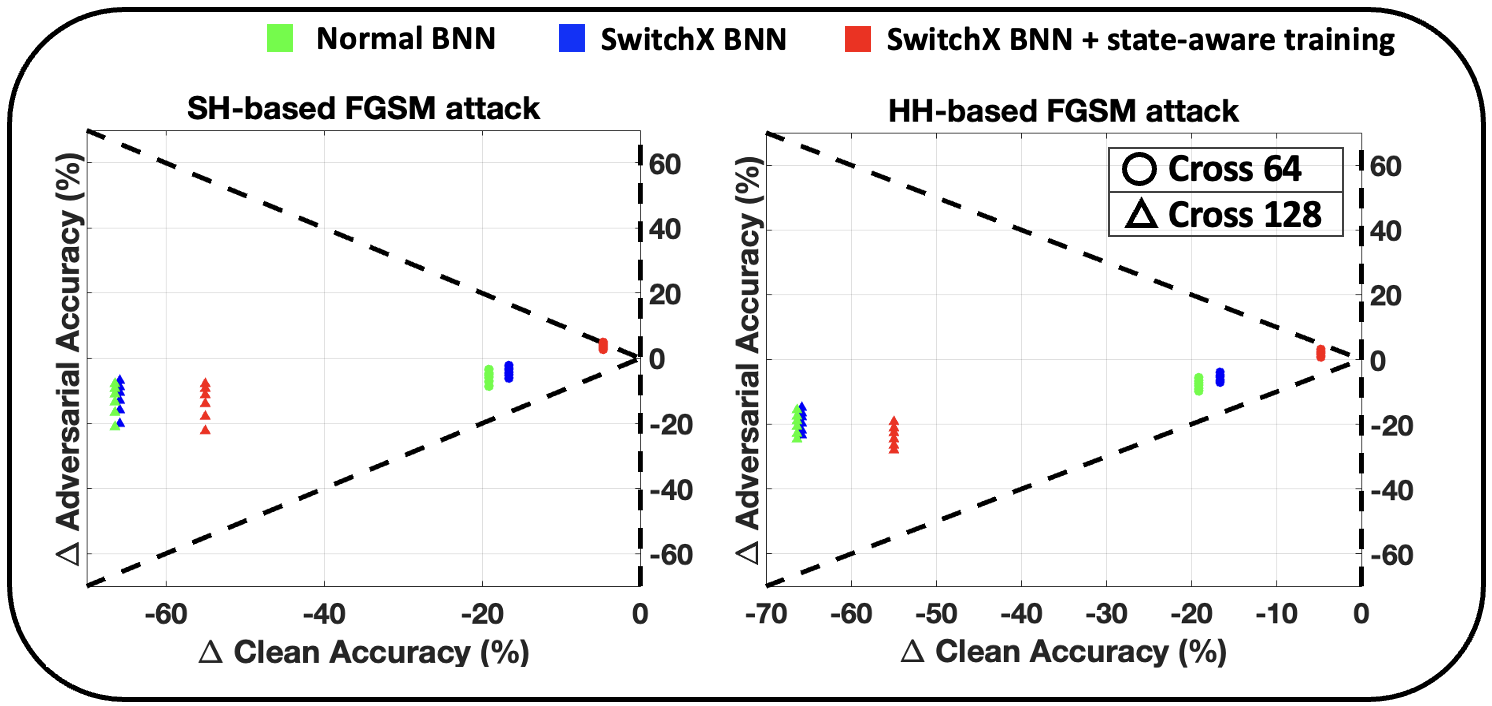}%
    \caption{Robustness maps for VGG16 BNN with CIFAR-10 dataset for SH (Left) and HH (Right) modes of FGSM attack respectively mapped onto 64$\times$64 (circle) and 128$\times$128 (triangle) crossbars for both standalone \textit{SwitchX} BNN and \textit{SwitchX} BNN + state-aware training}
    \label{64_128}
    
\end{figure}

\textbf{Effect of synaptic device variations on larger crossbars with greater $R_{MIN}$:}

Here, we increase $R_{MIN}$ of the NVM devices to $200 k\Omega$ (at ON/OFF ratio of 10) so that the impact of interconnect parasitic non-idealities can be minimized, thereby making simulations on larger crossbar sizes with a modest runtime feasible. In this scenario, the synaptic device variations become predominant and determine the robustness of the crossbar-mapped BNN models. For the robustness maps in Fig. \ref{64_128}, we assume the NVM device variations to be Gaussian with $\sigma/\mu = 20\%$ and the BNN mappings are carried out on 64$\times$64 and 128$\times$128 crossbars. The results are consistent with that in Fig. \ref{cifar10} (a) \& (b), indicating that the \textit{SwitchX} method is applicable to BNN mapped onto larger crossbar sizes. 

\subsection{Impact of \textit{SwitchX} on crossbar power consumption}

\label{switchx_power}

\begin{figure*}[t]
    \centering
    \includegraphics[width=\linewidth]{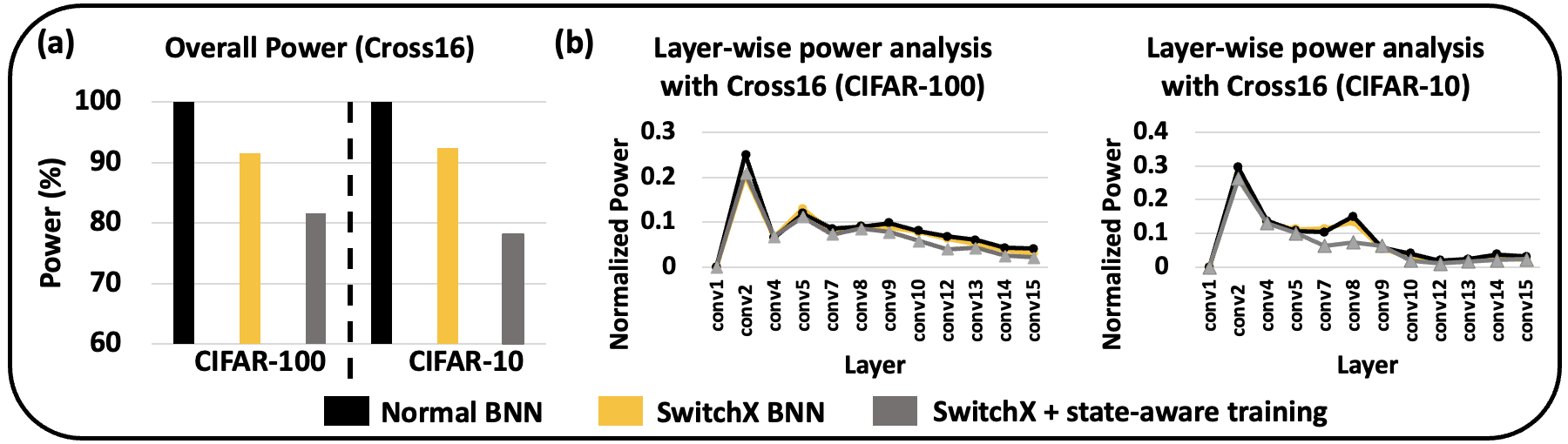}%
    \caption{(a) Plot showing power savings in 16$\times$16 crossbar for standalone \textit{SwitchX} as well as \textit{SwitchX} combined with state-aware training with respect to the baseline of Normal BNN for VGG16 BNN; (b) Plot showing layer-wise crossbar power consumption for the different convolutional layers in the VGG16 BNN using CIFAR-10 and CIFAR-100 datasets in case of mapping on 16$\times$16 crossbars}
    \label{power-cifar}
    
\end{figure*}

\begin{figure}[t]
    \centering
    \includegraphics[width=0.45\linewidth]{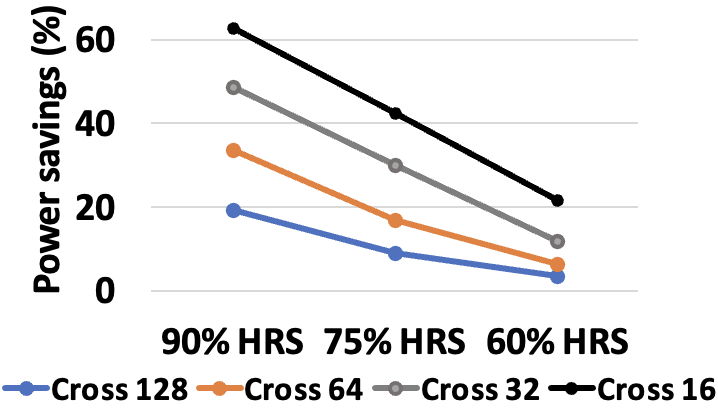}%
    \caption{Plot showing power savings on various crossbar sizes with \textit{SwitchX} mapping}
    \label{power}
    
\end{figure}

Owing to low leakage of the memristive devices, the $V.I$ power consumption in the crossbars constitutes a significantly large portion of the overall power expended in low-precision BNN inference on ReRAM crossbars \cite{sara, Chang2015, kim2018neural}. Thus, in this section we show an interesting by-product of \textit{SwitchX} mapping, whereby we can also reduce the crossbar power consumption compared to normally mapped BNNs (shown in Fig. \ref{power-cifar}). Here, the simulations for estimating the $V.I$ power consumed by the crossbars during BNN inference has been performed using ReRAM devices with $R_{MIN} = 20 k\Omega$ and ON/OFF ratio of 10 using the NeuroSim tool \cite{peng2019dnn+}, and the analog voltages input to the crossbar are +0.1/-0.1 V (obtained via SPICE simulations as specified in Section \ref{memxbar}). 

\textbf{Trends for energy-efficiency in crossbars: }\figurename{~\ref{power}} shows a plot of the power savings observed when randomly generated binary weight matrices with higher proportion of `+1' values are mapped via \textit{SwitchX} onto crossbars of sizes ranging from 16$\times$16, to 128$\times$128. Here, input voltages to the crossbars are drawn from a uniform distribution. The case of `90\% HRS states' implies that 10\% of the values in the BNN weight matrix were `-1', while `75\% HRS states' implies that 25\% of the values were `-1', \textit{i.e.} the weight matrix is less non-uniformly distributed than the former. Similarly, `60\% HRS states' implies that 40\% of the values in the BNN weight matrix were `-1'. We find that the power savings ($\sim 7-34\%$ on 64$\times$64 crossbars) increase when the distribution of BNN weights becomes more non-uniform (from `60\% HRS states' to `90\% HRS states') for different crossbar sizes. A more non-uniform distribution would imply greater proportion of HRS states on \textit{SwitchX} mapping, thereby translating into greater crossbar power savings owing to lower dot-product currents. Note, state-aware training in BNNs increases the non-uniformity in the distribution of HRS-LRS synapses in crossbars when mapped using \textit{SwitchX}. 

From \figurename{~\ref{power-cifar}}(a) (data shown for a 16$\times$16 crossbar), we find that standalone \textit{SwitchX} approach leads to $\sim9\%$ power-savings on an average with respect to Normal BNN for a VGG16 network with CIFAR-100 dataset. While for CIFAR-10 dataset, there is $\sim8\%$ power saving for standalone \textit{SwitchX}, it increases to $\sim22\%$ when \textit{SwitchX} is combined with state-aware training. Furthermore, on carrying out similar experiments on 32$\times$32 crossbars, we obtain $\sim21\%$ power savings on combining \textit{SwitchX} with state-aware training (data not shown for brevity). For CIFAR-100 dataset, we obtain $\sim19\%$ power savings on 16$\times$16 crossbars on combining \textit{SwitchX} with state-aware training. \figurename{~\ref{power-cifar}}(b) shows the layer-wise normalized average power consumption by the network with CIFAR-100 and CIFAR-10 dataset. We find that overall for each convolutional layer of the network, the power consumed by the \textit{SwitchX} BNN is lesser than the Normal BNN. Furthermore, this reduction becomes even more significant between `conv7' and `conv8' layers when \textit{SwitchX} is combined with state-aware training (for CIFAR-10 dataset). This result is in accordance with \figurename{~\ref{power}} which shows that the power savings on crossbar arrays increase significantly when the HRS-LRS state-distribution becomes more non-uniform.

\subsection{Comparison with previous works}
\label{sec:comp_switchx}

\begin{table}[t]
\caption{Table showing comparison with previous works of BNNs on crossbars. The improvements in energy-efficiency and adversarial accuracy (wherever applicable) have been quoted as is from the previous works. Here, N.A. stands for not applicable. }
\label{tab:comp_rel}
\begin{tabular}{|c|c|c|c|}
\hline
\textbf{Work} & \textbf{Type of non-ideality} & \textbf{Energy-efficiency} & \textbf{Adversarial Robustness (\%)} \\ \hline
REBNN \cite{song2019rebnn}                & N.A.          & 4.26$\times$          & N.A.          \\ \hline
Lattice \cite{zheng2020lattice}                & N.A.          & 7.68$\times$          & N.A.         \\ \hline
\cite{cherupally2022improving}                & Device noise          &  N.A.         & $\sim6\%~\uparrow$ (Black-box)          \\ \hline
\textbf{\textit{SwitchX}} & \textbf{Resistive \& Device noises} & \textbf{1.21-1.22$\times^*$} & \textbf{$\sim6-16\%~\uparrow$ (White-box)} \\ \hline
\multicolumn{2}{l}{\scriptsize$^*$Crossbar energy-efficiency\normalsize} & \multicolumn{1}{c}{}  & \multicolumn{1}{c}{} \\
\end{tabular}%
\end{table}

In Section \ref{related}, we have specified that there are several previous works on neural network implementations on RRAM crossbar-arrays, such as \cite{song2019rebnn, zheng2020lattice}, where the key goal is to obtain high energy-efficiency and not adversarial robustness. Although recently, \cite{cherupally2022improving} proposed a RRAM device noise-aware training methodology to generate BNN models that help improve natural as well as adversarial robustness on crossbars with device variations, it does not take into account the impact of circuit-level resistive non-idealities such as crossbar-interconnect parasitics. Also, the noise-aware training methodology in \cite{cherupally2022improving} does not translate to any benefits in terms of energy-efficiency for BNNs on crossbars. However, our \textit{SwitchX} mapping strategy has an interesting by-product of improving crossbar energy-efficiency in addition to providing better adversarial robustness (see Section \ref{switchx_power}). 
In Table \ref{tab:comp_rel}, we quantitatively compare our results with some of prior works on BNNs implemented on crossbars, including \cite{cherupally2022improving}, to put our work in context. Since the scope of this work is primarily to improve the adversarial robustness of BNNs on non-ideal crossbars, we do not achieve the best of energy-efficiency. However, we obtain significantly better robustness against stronger white-box adversarial attacks compared to \cite{cherupally2022improving} even upon the consideration of resistive crossbar non-idealities, which have a larger disruptive effect on the performance of neural networks.

\section{Conclusion}

In this work, we propose \textit{SwitchX}-mapping of binary weights onto crossbars in such a manner that a crossbar array always comprises of greater proportion of high resistance synapses (HRS) than low resistance synapses (LRS). We perform a comprehensive analysis on crossbar arrays comprising of interconnect and device-level non-idealities, whereby \textit{SwitchX} reduces the overall impact of non-idealities during inference. This effect manifests as reduced accuracy losses and increased robustness against adversarial attacks for BNNs mapped onto crossbars. We also combine \textit{SwitchX} with state-aware training that further increases the feasibility of HRS states during weight mapping to boost the adversarial robustness of BNNs on crossbars. With increase in the feasibility of HRS synapses on crossbars, \textit{SwitchX} also helps improve the crossbar energy-efficiencies, in addition to robustness.

\begin{acks}
This work was supported in part by C-BRIC, a JUMP center sponsored by DARPA and SRC, Google Research Scholar Award, the National Science Foundation (Grant \#1947826), TII (Abu Dhabi), the DARPA AI Exploration (AIE) program, and the DoE MMICC center SEA-CROGS (Award \#DE-SC0023198).
\end{acks}

\bibliographystyle{ACM-Reference-Format}
\bibliography{sample-base}


\begin{thebibliography}{57}


\ifx \showCODEN    \undefined \def \showCODEN     #1{\unskip}     \fi
\ifx \showDOI      \undefined \def \showDOI       #1{#1}\fi
\ifx \showISBNx    \undefined \def \showISBNx     #1{\unskip}     \fi
\ifx \showISBNxiii \undefined \def \showISBNxiii  #1{\unskip}     \fi
\ifx \showISSN     \undefined \def \showISSN      #1{\unskip}     \fi
\ifx \showLCCN     \undefined \def \showLCCN      #1{\unskip}     \fi
\ifx \shownote     \undefined \def \shownote      #1{#1}          \fi
\ifx \showarticletitle \undefined \def \showarticletitle #1{#1}   \fi
\ifx \showURL      \undefined \def \showURL       {\relax}        \fi
\providecommand\bibfield[2]{#2}
\providecommand\bibinfo[2]{#2}
\providecommand\natexlab[1]{#1}
\providecommand\showeprint[2][]{arXiv:#2}

\bibitem[Ankit et~al\mbox{.}(2019)]%
        {puma}
\bibfield{author}{\bibinfo{person}{Ankit} {et~al\mbox{.}}}
  \bibinfo{year}{2019}\natexlab{}.
\newblock \showarticletitle{PUMA: A Programmable Ultra-efficient
  Memristor-based Accelerator for Machine Learning Inference}.
\newblock \bibinfo{journal}{\emph{arXiv:1901.10351}} (\bibinfo{year}{2019}).
\newblock


\bibitem[Bhattacharjee et~al\mbox{.}(2020)]%
        {bhatta}
\bibfield{author}{\bibinfo{person}{Bhattacharjee} {et~al\mbox{.}}}
  \bibinfo{year}{2020}\natexlab{}.
\newblock \showarticletitle{Rethinking Non-idealities in Memristive Crossbars
  for Adversarial Robustness in Neural Networks}.
\newblock \bibinfo{journal}{\emph{arXiv:2008.11298}} (\bibinfo{year}{2020}).
\newblock


\bibitem[Bhattacharjee et~al\mbox{.}(2021a)]%
        {bhattacharjee2021efficiency}
\bibfield{author}{\bibinfo{person}{Bhattacharjee} {et~al\mbox{.}}}
  \bibinfo{year}{2021}\natexlab{a}.
\newblock \showarticletitle{Efficiency-driven Hardware Optimization for
  Adversarially Robust Neural Networks}.
\newblock \bibinfo{journal}{\emph{Design, Automation and Test in Europe
  Conference}} (\bibinfo{year}{2021}).
\newblock


\bibitem[Bhattacharjee et~al\mbox{.}(2021b)]%
        {neat}
\bibfield{author}{\bibinfo{person}{Abhiroop Bhattacharjee} {et~al\mbox{.}}}
  \bibinfo{year}{2021}\natexlab{b}.
\newblock \showarticletitle{NEAT: Non-linearity Aware Training for Accurate,
  Energy-Efficient and Robust Implementation of Neural Networks on 1T-1R
  Crossbars}.
\newblock \bibinfo{journal}{\emph{IEEE TCAD}} (\bibinfo{year}{2021}).
\newblock


\bibitem[Carlini et~al\mbox{.}(2019)]%
        {carlini}
\bibfield{author}{\bibinfo{person}{Carlini} {et~al\mbox{.}}}
  \bibinfo{year}{2019}\natexlab{}.
\newblock \showarticletitle{On Evaluating Adversarial Robustness}.
\newblock \bibinfo{journal}{\emph{arXiv:1902.06705}} (\bibinfo{year}{2019}).
\newblock


\bibitem[Chakraborty et~al\mbox{.}(2020a)]%
        {geniex}
\bibfield{author}{\bibinfo{person}{Chakraborty} {et~al\mbox{.}}}
  \bibinfo{year}{2020}\natexlab{a}.
\newblock \showarticletitle{GENIEx: A Generalized Approach to Emulating
  Non-Ideality in Memristive Xbars using Neural Networks}.
\newblock \bibinfo{journal}{\emph{arXiv:2003.06902}} (\bibinfo{year}{2020}).
\newblock


\bibitem[Chakraborty et~al\mbox{.}(2020b)]%
        {chakraborty2020pathways}
\bibfield{author}{\bibinfo{person}{Chakraborty} {et~al\mbox{.}}}
  \bibinfo{year}{2020}\natexlab{b}.
\newblock \showarticletitle{Pathways to efficient neuromorphic computing with
  non-volatile memory technologies}.
\newblock \bibinfo{journal}{\emph{Applied Physics Reviews}}
  \bibinfo{volume}{7}, \bibinfo{number}{2} (\bibinfo{year}{2020}),
  \bibinfo{pages}{021308}.
\newblock


\bibitem[Chang et~al\mbox{.}(2015)]%
        {Chang2015}
\bibfield{author}{\bibinfo{person}{Chang} {et~al\mbox{.}}}
  \bibinfo{year}{2015}\natexlab{}.
\newblock \showarticletitle{Challenges and Circuit Techniques for
  Energy-Efficient On-Chip Nonvolatile Memory Using Memristive Devices}.
\newblock \bibinfo{journal}{\emph{IEEE JETCAS}}  \bibinfo{volume}{5}
  (\bibinfo{year}{2015}), \bibinfo{pages}{183--193}.
\newblock


\bibitem[Chen et~al\mbox{.}(2017)]%
        {chen}
\bibfield{author}{\bibinfo{person}{Chen} {et~al\mbox{.}}}
  \bibinfo{year}{2017}\natexlab{}.
\newblock \showarticletitle{Circuit design for beyond von Neumann applications
  using emerging memory: From nonvolatile logics to neuromorphic computing}.
\newblock \bibinfo{journal}{\emph{ISQED}} (\bibinfo{year}{2017}),
  \bibinfo{pages}{23--28}.
\newblock


\bibitem[Chen et~al\mbox{.}(2015)]%
        {chen2015mitigating}
\bibfield{author}{\bibinfo{person}{Pai-Yu Chen}, \bibinfo{person}{Binbin Lin},
  \bibinfo{person}{I-Ting Wang}, \bibinfo{person}{Tuo-Hung Hou},
  \bibinfo{person}{Jieping Ye}, \bibinfo{person}{Sarma Vrudhula},
  \bibinfo{person}{Jae-sun Seo}, \bibinfo{person}{Yu Cao}, {and}
  \bibinfo{person}{Shimeng Yu}.} \bibinfo{year}{2015}\natexlab{}.
\newblock \showarticletitle{Mitigating effects of non-ideal synaptic device
  characteristics for on-chip learning}. In \bibinfo{booktitle}{\emph{2015
  IEEE/ACM International Conference on Computer-Aided Design (ICCAD)}}. IEEE,
  \bibinfo{pages}{194--199}.
\newblock


\bibitem[Chen et~al\mbox{.}(2018)]%
        {chen201865nm}
\bibfield{author}{\bibinfo{person}{Wei-Hao Chen}, \bibinfo{person}{Kai-Xiang
  Li}, \bibinfo{person}{Wei-Yu Lin}, \bibinfo{person}{Kuo-Hsiang Hsu},
  \bibinfo{person}{Pin-Yi Li}, \bibinfo{person}{Cheng-Han Yang},
  \bibinfo{person}{Cheng-Xin Xue}, \bibinfo{person}{En-Yu Yang},
  \bibinfo{person}{Yen-Kai Chen}, \bibinfo{person}{Yun-Sheng Chang},
  {et~al\mbox{.}}} \bibinfo{year}{2018}\natexlab{}.
\newblock \showarticletitle{A 65nm 1Mb nonvolatile computing-in-memory ReRAM
  macro with sub-16ns multiply-and-accumulate for binary DNN AI edge
  processors}. In \bibinfo{booktitle}{\emph{2018 IEEE International Solid-State
  Circuits Conference-(ISSCC)}}. IEEE, \bibinfo{pages}{494--496}.
\newblock


\bibitem[Cherupally et~al\mbox{.}(2022)]%
        {cherupally2022improving}
\bibfield{author}{\bibinfo{person}{Sai~Kiran Cherupally}, \bibinfo{person}{Jian
  Meng}, \bibinfo{person}{Adnan~Siraj Rakin}, \bibinfo{person}{Shihui Yin},
  \bibinfo{person}{Injune Yeo}, \bibinfo{person}{Shimeng Yu},
  \bibinfo{person}{Deliang Fan}, {and} \bibinfo{person}{Jae-Sun Seo}.}
  \bibinfo{year}{2022}\natexlab{}.
\newblock \showarticletitle{Improving the accuracy and robustness of RRAM-based
  in-memory computing against RRAM hardware noise and adversarial attacks}.
\newblock \bibinfo{journal}{\emph{Semiconductor Science and Technology}}
  \bibinfo{volume}{37}, \bibinfo{number}{3} (\bibinfo{year}{2022}),
  \bibinfo{pages}{034001}.
\newblock


\bibitem[Deng et~al\mbox{.}(2012)]%
        {deng2012rram}
\bibfield{author}{\bibinfo{person}{Yexin Deng}, \bibinfo{person}{Peng Huang},
  \bibinfo{person}{Bing Chen}, \bibinfo{person}{Xiaolin Yang},
  \bibinfo{person}{Bin Gao}, \bibinfo{person}{Juncheng Wang},
  \bibinfo{person}{Lang Zeng}, \bibinfo{person}{Gang Du},
  \bibinfo{person}{Jinfeng Kang}, {and} \bibinfo{person}{Xiaoyan Liu}.}
  \bibinfo{year}{2012}\natexlab{}.
\newblock \showarticletitle{RRAM crossbar array with cell selection device: A
  device and circuit interaction study}.
\newblock \bibinfo{journal}{\emph{IEEE transactions on Electron Devices}}
  \bibinfo{volume}{60}, \bibinfo{number}{2} (\bibinfo{year}{2012}),
  \bibinfo{pages}{719--726}.
\newblock


\bibitem[Fan et~al\mbox{.}(2015)]%
        {sttsnn}
\bibfield{author}{\bibinfo{person}{Fan} {et~al\mbox{.}}}
  \bibinfo{year}{2015}\natexlab{}.
\newblock \showarticletitle{STT-SNN: A Spin-Transfer-Torque Based Soft-Limiting
  Non-Linear Neuron for Low-Power Artificial Neural Networks}.
\newblock \bibinfo{journal}{\emph{IEEE Transactions on Nanotechnology}}
  \bibinfo{volume}{14}, \bibinfo{number}{6} (\bibinfo{year}{2015}),
  \bibinfo{pages}{1013--1023}.
\newblock


\bibitem[Goodfellow et~al\mbox{.}(2014)]%
        {goodfellow}
\bibfield{author}{\bibinfo{person}{Goodfellow} {et~al\mbox{.}}}
  \bibinfo{year}{2014}\natexlab{}.
\newblock \showarticletitle{Explaining and Harnessing Adversarial Examples}.
\newblock \bibinfo{journal}{\emph{arXiv:1412.6572}} (\bibinfo{year}{2014}).
\newblock


\bibitem[Gui et~al\mbox{.}(2019)]%
        {NEURIPS2019_2ca65f58}
\bibfield{author}{\bibinfo{person}{Shupeng Gui}, \bibinfo{person}{Haotao Wang},
  \bibinfo{person}{Haichuan Yang}, \bibinfo{person}{Chen Yu},
  \bibinfo{person}{Zhangyang Wang}, {and} \bibinfo{person}{Ji Liu}.}
  \bibinfo{year}{2019}\natexlab{}.
\newblock \showarticletitle{Model Compression with Adversarial Robustness: A
  Unified Optimization Framework}. In \bibinfo{booktitle}{\emph{Advances in
  Neural Information Processing Systems}},
  \bibfield{editor}{\bibinfo{person}{H.~Wallach},
  \bibinfo{person}{H.~Larochelle}, \bibinfo{person}{A.~Beygelzimer},
  \bibinfo{person}{F.~d\textquotesingle Alch\'{e}-Buc},
  \bibinfo{person}{E.~Fox}, {and} \bibinfo{person}{R.~Garnett}} (Eds.),
  Vol.~\bibinfo{volume}{32}. \bibinfo{publisher}{Curran Associates, Inc.}
\newblock
\urldef\tempurl%
\url{https://proceedings.neurips.cc/paper/2019/file/2ca65f58e35d9ad45bf7f3ae5cfd08f1-Paper.pdf}
\showURL{%
\tempurl}


\bibitem[Hajri et~al\mbox{.}(2019)]%
        {hajri2019rram}
\bibfield{author}{\bibinfo{person}{Basma Hajri}, \bibinfo{person}{Hassen
  Aziza}, \bibinfo{person}{Mohammad~M Mansour}, {and} \bibinfo{person}{Ali
  Chehab}.} \bibinfo{year}{2019}\natexlab{}.
\newblock \showarticletitle{RRAM device models: A comparative analysis with
  experimental validation}.
\newblock \bibinfo{journal}{\emph{IEEE Access}}  \bibinfo{volume}{7}
  (\bibinfo{year}{2019}), \bibinfo{pages}{168963--168980}.
\newblock


\bibitem[He et~al\mbox{.}(2020)]%
        {sara}
\bibfield{author}{\bibinfo{person}{He} {et~al\mbox{.}}}
  \bibinfo{year}{2020}\natexlab{}.
\newblock \showarticletitle{Towards State-Aware Computation in ReRAM Neural
  Networks}.
\newblock \bibinfo{journal}{\emph{Design and Automation Conference}}
  (\bibinfo{year}{2020}).
\newblock


\bibitem[He et~al\mbox{.}(2021)]%
        {sara_new}
\bibfield{author}{\bibinfo{person}{Yintao He}, \bibinfo{person}{Ying Wang},
  \bibinfo{person}{Huawei Li}, {and} \bibinfo{person}{Xiaowei Li}.}
  \bibinfo{year}{2021}\natexlab{}.
\newblock \showarticletitle{Saving Energy of RRAM-based Neural Accelerator
  through State-Aware Computing}.
\newblock \bibinfo{journal}{\emph{IEEE Transactions on Computer-Aided Design of
  Integrated Circuits and Systems}} (\bibinfo{year}{2021}),
  \bibinfo{pages}{1--1}.
\newblock
\urldef\tempurl%
\url{https://doi.org/10.1109/TCAD.2021.3103147}
\showDOI{\tempurl}


\bibitem[Hirtzlin et~al\mbox{.}(2019)]%
        {hirtzlin2019outstanding}
\bibfield{author}{\bibinfo{person}{Tifenn Hirtzlin}, \bibinfo{person}{Marc
  Bocquet}, \bibinfo{person}{J-O Klein}, \bibinfo{person}{Etienne Nowak},
  \bibinfo{person}{Elisa Vianello}, \bibinfo{person}{J-M Portal}, {and}
  \bibinfo{person}{Damien Querlioz}.} \bibinfo{year}{2019}\natexlab{}.
\newblock \showarticletitle{Outstanding bit error tolerance of resistive
  ram-based binarized neural networks}. In \bibinfo{booktitle}{\emph{2019 IEEE
  International Conference on Artificial Intelligence Circuits and Systems
  (AICAS)}}. IEEE, \bibinfo{pages}{288--292}.
\newblock


\bibitem[Hubara et~al\mbox{.}(2016)]%
        {hubara2016binarized}
\bibfield{author}{\bibinfo{person}{Itay Hubara}, \bibinfo{person}{Matthieu
  Courbariaux}, \bibinfo{person}{Daniel Soudry}, \bibinfo{person}{Ran
  El-Yaniv}, {and} \bibinfo{person}{Yoshua Bengio}.}
  \bibinfo{year}{2016}\natexlab{}.
\newblock \showarticletitle{Binarized neural networks}.
\newblock \bibinfo{journal}{\emph{Advances in neural information processing
  systems}}  \bibinfo{volume}{29} (\bibinfo{year}{2016}).
\newblock


\bibitem[Imani et~al\mbox{.}(2019)]%
        {imani2019floatpim}
\bibfield{author}{\bibinfo{person}{Mohsen Imani}, \bibinfo{person}{Saransh
  Gupta}, \bibinfo{person}{Yeseong Kim}, {and} \bibinfo{person}{Tajana
  Rosing}.} \bibinfo{year}{2019}\natexlab{}.
\newblock \showarticletitle{Floatpim: In-memory acceleration of deep neural
  network training with high precision}. In \bibinfo{booktitle}{\emph{2019
  ACM/IEEE 46th Annual International Symposium on Computer Architecture
  (ISCA)}}. IEEE, \bibinfo{pages}{802--815}.
\newblock


\bibitem[Jain et~al\mbox{.}(2018)]%
        {rxnn}
\bibfield{author}{\bibinfo{person}{Jain} {et~al\mbox{.}}}
  \bibinfo{year}{2018}\natexlab{}.
\newblock \showarticletitle{RxNN: A Framework for Evaluating Deep Neural
  Networks on Resistive Crossbars}.
\newblock \bibinfo{journal}{\emph{arXiv:1809.00072}} (\bibinfo{year}{2018}).
\newblock


\bibitem[Jain et~al\mbox{.}(2019)]%
        {cxdnn}
\bibfield{author}{\bibinfo{person}{Jain} {et~al\mbox{.}}}
  \bibinfo{year}{2019}\natexlab{}.
\newblock \showarticletitle{CxDNN: Hardware-Software Compensation Methods for
  Deep Neural Networks on Resistive Crossbar Systems}.
\newblock \bibinfo{journal}{\emph{ACM Trans. Embed. Comput. Syst.}}
  \bibinfo{volume}{18}, \bibinfo{number}{6}, Article \bibinfo{articleno}{113}
  (\bibinfo{date}{Nov.} \bibinfo{year}{2019}), \bibinfo{numpages}{23}~pages.
\newblock
\showISSN{1539-9087}
\urldef\tempurl%
\url{https://doi.org/10.1145/3362035}
\showDOI{\tempurl}


\bibitem[Jiang et~al\mbox{.}(2014)]%
        {nanoHUB.org19}
\bibfield{author}{\bibinfo{person}{Jiang} {et~al\mbox{.}}}
  \bibinfo{year}{2014}\natexlab{}.
\newblock \bibinfo{title}{Stanford University Resistive-Switching Random Access
  Memory (RRAM) Verilog-A Model}.
\newblock
\newblock
\urldef\tempurl%
\url{https://doi.org/doi:/10.4231/D37H1DN48}
\showDOI{\tempurl}


\bibitem[Kim et~al\mbox{.}(2019)]%
        {membin1}
\bibfield{author}{\bibinfo{person}{Kim} {et~al\mbox{.}}}
  \bibinfo{year}{2019}\natexlab{}.
\newblock \showarticletitle{Memristor crossbar array for binarized neural
  networks}.
\newblock \bibinfo{journal}{\emph{AIP Advances}} \bibinfo{volume}{9},
  \bibinfo{number}{4} (\bibinfo{year}{2019}), \bibinfo{pages}{045131}.
\newblock
\urldef\tempurl%
\url{https://doi.org/10.1063/1.5092177}
\showDOI{\tempurl}
\showeprint{https://doi.org/10.1063/1.5092177}


\bibitem[Kim et~al\mbox{.}(2018)]%
        {kim2018neural}
\bibfield{author}{\bibinfo{person}{Yulhwa Kim}, \bibinfo{person}{Hyungjun Kim},
  {and} \bibinfo{person}{Jae-Joon Kim}.} \bibinfo{year}{2018}\natexlab{}.
\newblock \showarticletitle{Neural network-hardware co-design for scalable
  RRAM-based BNN accelerators}.
\newblock \bibinfo{journal}{\emph{arXiv preprint arXiv:1811.02187}}
  (\bibinfo{year}{2018}).
\newblock


\bibitem[Krizhevsky(2009)]%
        {cifar}
\bibfield{author}{\bibinfo{person}{Alex Krizhevsky}.}
  \bibinfo{year}{2009}\natexlab{}.
\newblock \bibinfo{booktitle}{\emph{Learning multiple layers of features from
  tiny images}}.
\newblock \bibinfo{type}{{T}echnical {R}eport}.
\newblock


\bibitem[Kurakin et~al\mbox{.}(2016)]%
        {ensat}
\bibfield{author}{\bibinfo{person}{Kurakin} {et~al\mbox{.}}}
  \bibinfo{year}{2016}\natexlab{}.
\newblock \showarticletitle{Adversarial Machine Learning at Scale}.
\newblock \bibinfo{journal}{\emph{arXiv:1611.01236}} (\bibinfo{year}{2016}).
\newblock


\bibitem[Lee et~al\mbox{.}(2017)]%
        {gat}
\bibfield{author}{\bibinfo{person}{Lee}, \bibinfo{person}{}, {et~al\mbox{.}}}
  \bibinfo{year}{2017}\natexlab{}.
\newblock \showarticletitle{Generative Adversarial Trainer: Defense to
  Adversarial Perturbations with GAN}.
\newblock \bibinfo{journal}{\emph{arXiv:1705.03387}} (\bibinfo{year}{2017}).
\newblock


\bibitem[Lin et~al\mbox{.}(2019)]%
        {defquant}
\bibfield{author}{\bibinfo{person}{Lin} {et~al\mbox{.}}}
  \bibinfo{year}{2019}\natexlab{}.
\newblock \showarticletitle{Defensive Quantization: When Efficiency Meets
  Robustness}.
\newblock \bibinfo{journal}{\emph{arXiv:1904.08444}} (\bibinfo{year}{2019}).
\newblock


\bibitem[Liu et~al\mbox{.}(2014)]%
        {liu}
\bibfield{author}{\bibinfo{person}{Liu} {et~al\mbox{.}}}
  \bibinfo{year}{2014}\natexlab{}.
\newblock \showarticletitle{Reduction and IR-drop compensations techniques for
  reliable neuromorphic computing systems}.
\newblock \bibinfo{journal}{\emph{2014 IEEE/ACM International Conference on
  Computer-Aided Design (ICCAD)}} (\bibinfo{year}{2014}),
  \bibinfo{pages}{63--70}.
\newblock


\bibitem[Liu et~al\mbox{.}(2015)]%
        {vortex}
\bibfield{author}{\bibinfo{person}{Beiye Liu}, \bibinfo{person}{Hai Li},
  \bibinfo{person}{Yiran Chen}, \bibinfo{person}{Xin Li}, \bibinfo{person}{Qing
  Wu}, {and} \bibinfo{person}{Tingwen Huang}.} \bibinfo{year}{2015}\natexlab{}.
\newblock \showarticletitle{Vortex: variation-aware training for memristor
  x-bar}. In \bibinfo{booktitle}{\emph{Proceedings of the 52nd Annual Design
  Automation Conference}}. \bibinfo{pages}{1--6}.
\newblock


\bibitem[Madry et~al\mbox{.}(2017)]%
        {pgd}
\bibfield{author}{\bibinfo{person}{Madry} {et~al\mbox{.}}}
  \bibinfo{year}{2017}\natexlab{}.
\newblock \showarticletitle{Towards Deep Learning Models Resistant to
  Adversarial Attacks}.
\newblock \bibinfo{journal}{\emph{arXiv:1706.06083}} (\bibinfo{year}{2017}).
\newblock


\bibitem[Mehonic et~al\mbox{.}(2019)]%
        {mehonic2019simulation}
\bibfield{author}{\bibinfo{person}{Adnan Mehonic}, \bibinfo{person}{Dovydas
  Joksas}, \bibinfo{person}{Wing~H Ng}, \bibinfo{person}{Mark Buckwell}, {and}
  \bibinfo{person}{Anthony~J Kenyon}.} \bibinfo{year}{2019}\natexlab{}.
\newblock \showarticletitle{Simulation of inference accuracy using realistic
  RRAM devices}.
\newblock \bibinfo{journal}{\emph{Frontiers in neuroscience}}
  \bibinfo{volume}{13} (\bibinfo{year}{2019}), \bibinfo{pages}{593}.
\newblock


\bibitem[Ni et~al\mbox{.}(2017)]%
        {ni2017energy}
\bibfield{author}{\bibinfo{person}{Leibin Ni}, \bibinfo{person}{Zichuan Liu},
  \bibinfo{person}{Hao Yu}, {and} \bibinfo{person}{Rajiv~V Joshi}.}
  \bibinfo{year}{2017}\natexlab{}.
\newblock \showarticletitle{An energy-efficient digital ReRAM-crossbar-based
  CNN with bitwise parallelism}.
\newblock \bibinfo{journal}{\emph{IEEE Journal on Exploratory solid-state
  computational devices and circuits}}  \bibinfo{volume}{3}
  (\bibinfo{year}{2017}), \bibinfo{pages}{37--46}.
\newblock


\bibitem[Nie et~al\mbox{.}(2022)]%
        {nie2022cross}
\bibfield{author}{\bibinfo{person}{Chen Nie}, \bibinfo{person}{Zongwu Wang},
  \bibinfo{person}{Qidong Tang}, \bibinfo{person}{Chenyang Lv},
  \bibinfo{person}{Li Jiang}, {and} \bibinfo{person}{Zhezhi He}.}
  \bibinfo{year}{2022}\natexlab{}.
\newblock \showarticletitle{Cross-layer Designs against Non-ideal Effects in
  ReRAM-based Processing-in-Memory System}. In \bibinfo{booktitle}{\emph{2022
  23rd International Symposium on Quality Electronic Design (ISQED)}}. IEEE,
  \bibinfo{pages}{1--6}.
\newblock


\bibitem[Panda et~al\mbox{.}(2019)]%
        {pixeld}
\bibfield{author}{\bibinfo{person}{Panda} {et~al\mbox{.}}}
  \bibinfo{year}{2019}\natexlab{}.
\newblock \showarticletitle{Discretization Based Solutions for Secure Machine
  Learning Against Adversarial Attacks}.
\newblock \bibinfo{journal}{\emph{IEEE Access}}  \bibinfo{volume}{7}
  (\bibinfo{year}{2019}), \bibinfo{pages}{70157–70168}.
\newblock
\showISSN{2169-3536}
\urldef\tempurl%
\url{https://doi.org/10.1109/access.2019.2919463}
\showDOI{\tempurl}


\bibitem[Panda(2020)]%
        {quanos}
\bibfield{author}{\bibinfo{person}{Priyadarshini Panda}.}
  \bibinfo{year}{2020}\natexlab{}.
\newblock \showarticletitle{QUANOS: Adversarial Noise Sensitivity Driven Hybrid
  Quantization of Neural Networks}. In \bibinfo{booktitle}{\emph{ACM/IEEE
  ISLPED}} \emph{(\bibinfo{series}{ISLPED '20})}. \bibinfo{pages}{187–192}.
\newblock
\urldef\tempurl%
\url{https://doi.org/10.1145/3370748.3406585}
\showDOI{\tempurl}


\bibitem[Papernot et~al\mbox{.}(2016)]%
        {gradmask}
\bibfield{author}{\bibinfo{person}{Papernot} {et~al\mbox{.}}}
  \bibinfo{year}{2016}\natexlab{}.
\newblock \showarticletitle{Practical Black-Box Attacks against Machine
  Learning}.
\newblock \bibinfo{journal}{\emph{arXiv:1602.02697}} (\bibinfo{year}{2016}).
\newblock


\bibitem[Peng et~al\mbox{.}(2019)]%
        {peng2019dnn+}
\bibfield{author}{\bibinfo{person}{Xiaochen Peng}, \bibinfo{person}{Shanshi
  Huang}, \bibinfo{person}{Yandong Luo}, \bibinfo{person}{Xiaoyu Sun}, {and}
  \bibinfo{person}{Shimeng Yu}.} \bibinfo{year}{2019}\natexlab{}.
\newblock \showarticletitle{DNN+ NeuroSim: An end-to-end benchmarking framework
  for compute-in-memory accelerators with versatile device technologies}. In
  \bibinfo{booktitle}{\emph{2019 IEEE international electron devices meeting
  (IEDM)}}. IEEE, \bibinfo{pages}{32--5}.
\newblock


\bibitem[Roy et~al\mbox{.}(2020)]%
        {roy2020robustness}
\bibfield{author}{\bibinfo{person}{Deboleena Roy}, \bibinfo{person}{Indranil
  Chakraborty}, \bibinfo{person}{Timur Ibrayev}, {and} \bibinfo{person}{Kaushik
  Roy}.} \bibinfo{year}{2020}\natexlab{}.
\newblock \showarticletitle{Robustness Hidden in Plain Sight: Can Analog
  Computing Defend Against Adversarial Attacks?}
\newblock \bibinfo{journal}{\emph{arXiv: 2008.1201}} (\bibinfo{year}{2020}).
\newblock


\bibitem[Roy et~al\mbox{.}(2021)]%
        {roy2021intrinsic}
\bibfield{author}{\bibinfo{person}{Deboleena Roy}, \bibinfo{person}{Indranil
  Chakraborty}, \bibinfo{person}{Timur Ibrayev}, {and} \bibinfo{person}{Kaushik
  Roy}.} \bibinfo{year}{2021}\natexlab{}.
\newblock \showarticletitle{On the Intrinsic Robustness of NVM Crossbars
  Against Adversarial Attacks}. In \bibinfo{booktitle}{\emph{2021 58th ACM/IEEE
  Design Automation Conference (DAC)}}. IEEE, \bibinfo{pages}{565--570}.
\newblock


\bibitem[Schuman et~al\mbox{.}(2017)]%
        {schuman}
\bibfield{author}{\bibinfo{person}{Schuman} {et~al\mbox{.}}}
  \bibinfo{year}{2017}\natexlab{}.
\newblock \showarticletitle{A Survey of Neuromorphic Computing and Neural
  Networks in Hardware}.
\newblock \bibinfo{journal}{\emph{arXiv:1705.06963}} (\bibinfo{year}{2017}).
\newblock


\bibitem[Sehwag et~al\mbox{.}(2020)]%
        {sehwag2020hydra}
\bibfield{author}{\bibinfo{person}{Vikash Sehwag}, \bibinfo{person}{Shiqi
  Wang}, \bibinfo{person}{Prateek Mittal}, {and} \bibinfo{person}{Suman Jana}.}
  \bibinfo{year}{2020}\natexlab{}.
\newblock \showarticletitle{Hydra: Pruning adversarially robust neural
  networks}.
\newblock \bibinfo{journal}{\emph{Advances in Neural Information Processing
  Systems}}  \bibinfo{volume}{33} (\bibinfo{year}{2020}),
  \bibinfo{pages}{19655--19666}.
\newblock


\bibitem[Sengupta et~al\mbox{.}(2016)]%
        {sengupta}
\bibfield{author}{\bibinfo{person}{Sengupta} {et~al\mbox{.}}}
  \bibinfo{year}{2016}\natexlab{}.
\newblock \showarticletitle{Proposal for an All-Spin Artificial Neural Network:
  Emulating Neural and Synaptic Functionalities Through Domain Wall Motion in
  Ferromagnets}.
\newblock \bibinfo{journal}{\emph{IEEE TBCAS}} (\bibinfo{year}{2016}).
\newblock


\bibitem[Shafiee et~al\mbox{.}(2016)]%
        {shafiee2016isaac}
\bibfield{author}{\bibinfo{person}{Shafiee} {et~al\mbox{.}}}
  \bibinfo{year}{2016}\natexlab{}.
\newblock \showarticletitle{ISAAC: A convolutional neural network accelerator
  with in-situ analog arithmetic in crossbars}.
\newblock \bibinfo{journal}{\emph{ACM SIGARCH Computer Architecture News}}
  \bibinfo{volume}{44}, \bibinfo{number}{3} (\bibinfo{year}{2016}),
  \bibinfo{pages}{14--26}.
\newblock


\bibitem[Sharad et~al\mbox{.}(2012)]%
        {sharad}
\bibfield{author}{\bibinfo{person}{Sharad} {et~al\mbox{.}}}
  \bibinfo{year}{2012}\natexlab{}.
\newblock \showarticletitle{Spin neuron for ultra low power computational
  hardware}.
\newblock \bibinfo{journal}{\emph{70th Device Research Conference}}
  (\bibinfo{year}{2012}), \bibinfo{pages}{221--222}.
\newblock


\bibitem[Simonyan et~al\mbox{.}(2014)]%
        {vgg}
\bibfield{author}{\bibinfo{person}{Simonyan} {et~al\mbox{.}}}
  \bibinfo{year}{2014}\natexlab{}.
\newblock \showarticletitle{Very Deep Convolutional Networks for Large-Scale
  Image Recognition}.
\newblock \bibinfo{journal}{\emph{arXiv:1409.1556}} (\bibinfo{year}{2014}).
\newblock


\bibitem[Song et~al\mbox{.}(2017)]%
        {song2017pipelayer}
\bibfield{author}{\bibinfo{person}{Linghao Song}, \bibinfo{person}{Xuehai
  Qian}, \bibinfo{person}{Hai Li}, {and} \bibinfo{person}{Yiran Chen}.}
  \bibinfo{year}{2017}\natexlab{}.
\newblock \showarticletitle{Pipelayer: A pipelined reram-based accelerator for
  deep learning}. In \bibinfo{booktitle}{\emph{2017 IEEE international
  symposium on high performance computer architecture (HPCA)}}. IEEE,
  \bibinfo{pages}{541--552}.
\newblock


\bibitem[Song et~al\mbox{.}(2019)]%
        {song2019rebnn}
\bibfield{author}{\bibinfo{person}{Linghao Song}, \bibinfo{person}{You Wu},
  \bibinfo{person}{Xuehai Qian}, \bibinfo{person}{Hai Li}, {and}
  \bibinfo{person}{Yiran Chen}.} \bibinfo{year}{2019}\natexlab{}.
\newblock \showarticletitle{Rebnn: in-situ acceleration of binarized neural
  networks in reram using complementary resistive cell}.
\newblock \bibinfo{journal}{\emph{CCF Transactions on High Performance
  Computing}} \bibinfo{volume}{1}, \bibinfo{number}{3} (\bibinfo{year}{2019}),
  \bibinfo{pages}{196--208}.
\newblock


\bibitem[Tang et~al\mbox{.}(2022)]%
        {tang2022hawis}
\bibfield{author}{\bibinfo{person}{Qidong Tang}, \bibinfo{person}{Zhezhi He},
  \bibinfo{person}{Fangxin Liu}, \bibinfo{person}{Zongwu Wang},
  \bibinfo{person}{Yiyuan Zhou}, \bibinfo{person}{Yinghuan Zhang}, {and}
  \bibinfo{person}{Li Jiang}.} \bibinfo{year}{2022}\natexlab{}.
\newblock \showarticletitle{HAWIS: Hardware-Aware Automated WIdth Search for
  Accurate, Energy-Efficient and Robust Binary Neural Network on ReRAM
  Dot-Product Engine}. In \bibinfo{booktitle}{\emph{2022 27th Asia and South
  Pacific Design Automation Conference (ASP-DAC)}}. IEEE,
  \bibinfo{pages}{226--231}.
\newblock


\bibitem[Truong et~al\mbox{.}(2014)]%
        {membin2}
\bibfield{author}{\bibinfo{person}{Truong} {et~al\mbox{.}}}
  \bibinfo{year}{2014}\natexlab{}.
\newblock \showarticletitle{New Memristor-Based Crossbar Array Architecture
  with $50\%$ Area Reduction and $48\%$ Power Saving for Matrix-Vector
  Multiplication of Analog Neuromorphic Computing}.
\newblock \bibinfo{journal}{\emph{Journal of Semiconductor Technology and
  Science}}  \bibinfo{volume}{14} (\bibinfo{year}{2014}),
  \bibinfo{pages}{356--363}.
\newblock


\bibitem[Wong et~al\mbox{.}(2012)]%
        {mo-rram}
\bibfield{author}{\bibinfo{person}{Wong} {et~al\mbox{.}}}
  \bibinfo{year}{2012}\natexlab{}.
\newblock \showarticletitle{Metal–Oxide RRAM}.
\newblock \bibinfo{journal}{\emph{Proc. IEEE}} \bibinfo{volume}{100},
  \bibinfo{number}{6} (\bibinfo{year}{2012}), \bibinfo{pages}{1951--1970}.
\newblock


\bibitem[Xia et~al\mbox{.}(2016)]%
        {xia2016technological}
\bibfield{author}{\bibinfo{person}{Lixue Xia}, \bibinfo{person}{Peng Gu},
  \bibinfo{person}{Boxun Li}, \bibinfo{person}{Tianqi Tang},
  \bibinfo{person}{Xiling Yin}, \bibinfo{person}{Wenqin Huangfu},
  \bibinfo{person}{Shimeng Yu}, \bibinfo{person}{Yu Cao}, \bibinfo{person}{Yu
  Wang}, {and} \bibinfo{person}{Huazhong Yang}.}
  \bibinfo{year}{2016}\natexlab{}.
\newblock \showarticletitle{Technological exploration of RRAM crossbar array
  for matrix-vector multiplication}.
\newblock \bibinfo{journal}{\emph{Journal of Computer Science and Technology}}
  \bibinfo{volume}{31}, \bibinfo{number}{1} (\bibinfo{year}{2016}),
  \bibinfo{pages}{3--19}.
\newblock


\bibitem[Zhang et~al\mbox{.}(2020)]%
        {zhang2019handling}
\bibfield{author}{\bibinfo{person}{Baogang Zhang}, \bibinfo{person}{Necati
  Uysal}, \bibinfo{person}{Deliang Fan}, {and} \bibinfo{person}{Rickard
  Ewetz}.} \bibinfo{year}{2020}\natexlab{}.
\newblock \showarticletitle{Handling Stuck-at-Fault Defects Using Matrix
  Transformation for Robust Inference of DNNs}.
\newblock \bibinfo{journal}{\emph{IEEE Transactions on Computer-Aided Design of
  Integrated Circuits and Systems}} \bibinfo{volume}{39}, \bibinfo{number}{10}
  (\bibinfo{year}{2020}), \bibinfo{pages}{2448--2460}.
\newblock
\urldef\tempurl%
\url{https://doi.org/10.1109/TCAD.2019.2944582}
\showDOI{\tempurl}


\bibitem[Zheng et~al\mbox{.}(2020)]%
        {zheng2020lattice}
\bibfield{author}{\bibinfo{person}{Qilin Zheng}, \bibinfo{person}{Zongwei
  Wang}, \bibinfo{person}{Zishun Feng}, \bibinfo{person}{Bonan Yan},
  \bibinfo{person}{Yimao Cai}, \bibinfo{person}{Ru Huang},
  \bibinfo{person}{Yiran Chen}, \bibinfo{person}{Chia-Lin Yang}, {and}
  \bibinfo{person}{Hai~Helen Li}.} \bibinfo{year}{2020}\natexlab{}.
\newblock \showarticletitle{Lattice: an ADC/DAC-less ReRAM-based
  processing-in-memory architecture for accelerating deep convolution neural
  networks}. In \bibinfo{booktitle}{\emph{2020 57th ACM/IEEE Design Automation
  Conference (DAC)}}. IEEE, \bibinfo{pages}{1--6}.
\newblock


\end{thebibliography}


\end{document}